\documentclass[12pt,a4paper]{article}
\pdfoutput=1
\usepackage{lipsum}
\usepackage{authblk}
\usepackage[top=2cm, bottom=2cm, left=2cm, right=2cm]{geometry}
\usepackage{graphicx}
\usepackage{lineno}
\usepackage{algorithm2e}

\usepackage{placeins}

\newcommand{\indaco}[1]{\textcolor{black}{#1}}

\renewenvironment{abstract}{%
\hfill\begin{minipage}{0.95\textwidth}
\rule{\textwidth}{1pt}}
{\par\noindent\rule{\textwidth}{1pt}\end{minipage}}
\makeatletter
\renewcommand\@maketitle{%
\hfill
\begin{minipage}{0.95\textwidth}
\vskip 2em
\let\footnote\thanks 
{\LARGE \@title \par }
\vskip 1.5em
{\large \@author \par}
\end{minipage}
\vskip 1em \par
}
\makeatother

\usepackage{nameref,hyperref}
\usepackage{xcolor}

\begin{document}

\title{General scores for accessibility and inequality measures in urban areas}

\author[1,*]{Indaco Biazzo}
\author[2]{Bernardo Monechi}
\author[2,3,4]{Vittorio Loreto}

\affil[1]{Politecnico di Torino, Corso Duca degli Abruzzi 24, Torino, Italy}
\affil[2]{SONY Computer Science Laboratories, Paris, 6, rue Amyot, 75005, Paris, France}
\affil[3]{Complexity Science Hub, Josefst\"adter Strasse 39, A 1080 Vienna, Austria}
\affil[4]{Sapienza University of Rome, Physics Department, Piazzale Aldo Moro 2, 00185 Rome, Italy}

\affil[*]{indaco.biazzo@polito.it}

\flushbottom
\maketitle

\begin{abstract}
In the last decades, the acceleration of urban growth has led to an unprecedented level of urban interactions and interdependence.  
This situation calls for a significant effort among the scientific community to come up with engaging and meaningful visualizations and accessible scenario simulation engines. The present paper gives a contribution in this direction by providing general methods to evaluate accessibility in cities based on public transportation data. Through the notion of isochrones, \indaco{the accessibility quantities proposed measure the performance of transport systems at connecting places and people in urban systems.  Then} we introduce scores rank cities according to their overall accessibility. We highlight significant inequalities in the distribution \indaco{of these measures} across the population, which are found to be strikingly similar across various urban environments. 
Our results are released through the interactive platform: www.citychrone.org, aimed at providing the community at large with a useful tool for awareness and decision-making.
\end{abstract}

\thispagestyle{empty}
\section*{Keywords}
Accessibility, Science of city,  Isochrone, Public Transports, Urban Planning
\section*{Introduction}

The inherent complexity of the emerging challenges human beings collectively face requires a deep comprehension of the underlying phenomena in order to plan effective strategies and sustainable solutions.
Cties stand as a paramount example of how a complex interplay of infrastructures, technologies, and human behaviors may lead to outcomes and patterns very far from the usual cause-effect scheme~\cite{nations2014world}. The science of cities is a research area that greatly benefited from the digital revolution in the last decades~\cite{batty2013new}. Nowadays, the deployment of Information and Communication Technologies~\cite{mayer2013big} and the consequent availability of an unprecedented wealth of data is opening new opportunities for a scientific investigation into the complexity of urban environments.  This availability of data fostered studies aimed at identifying the patterns of co-evolution of human and social behaviors~ \cite{Gallotti2016,Alessandretti160156,1367-2630-5-1-348,mastroianni2015local} as well as innovation at the level of infrastructures and services~\cite{fleurquin2014characterization,Gallotti2015,monechi2015congestion,sen2003small,guimera2005worldwide}. This paper aims at contributing to the ongoing debate about the future of our cities and the way to combine growth~\cite{WorldUrbanization_2014} with efficiency and inclusiveness. To this end, we focus on a specific aspect of cities, namely the topic of accessibility. 
Accessibility can be described as the capacity of cities to allow people to move efficiently by guaranteeing equity and equal access to personal and professional opportunities. From this perspective, accessibility does not mean only the overall capacity of urban transit: it also needs to be inflected as the accessibility of specific areas, for particular people with specific purposes. It is not rare that public transport projects spearheaded by governments benefit only a tiny fraction of the population, and whilst the average traveling conditions remain poor for the majority of the population. It is thus important to be able to quantify accessibility in a way that closely represents the experience of citizens. 
Following a common approach in accessibility studies\cite{geurs2004accessibility}, we focus on traveling times between geographical areas which better represents the mindset citizens adopt in planning their mobility.
The key mathematical notion used to quantify traveling times will be that of {\em isochronic maps}, i.e. maps showing areas related to isochrones between different points. Considering a geographical point, its isochronic map will be composed by isochronic contours marking regions reachable in a given time-span, using different transportation systems. Isochronic maps exist since 1881, when Sir Francis Galton published the first isochronic map in the Proceedings of the Royal Geographical Society~\cite{galton1881construction}, showing travel times in days from London to different parts of the world. Nowadays, the availability of data related to mobility allows for the compilation of very accurate isochronic maps for different locations, different geographical areas, different social communities, and different transportation systems. 

Though the notion of isochrone is well defined, its computation depends on the transportation system adopted. Here we focus on public transportation, and we compute traveling times and isochrones using a routing approach that exploits multi-modality. This implies that the best route between two two points A and B in the city can be realized through a combination of several transportation means (walking, buses, metro lines, trains). For the sake of simplicity and without loss of generality, here we only consider the official public transport schedules for many cities in North America, Europe, and Australia. Following a recent interesting trend in scientific research~\cite{zastrow2015data}, we developed visualizations on maps of this body of information, as well as several metrics for accessibility, through the open CityChrone platform (http://www.citychrone.org). Data about real-time passages of public transport journeys or other public or private means of transport can be easily integrated into the platform as well.

Usually, studies about public transport analyze the networks of transport as static graphs, where the nodes represent stops and the edges represent the routes connecting them~\cite{Banavar1999,Louf2014, Masucci2015, Arcaute2016, Li2017}. Very few studies have instead incorporated in a systematic way the ``temporal" features of these systems~\cite{Alessandretti160156, Gallotti2015, boussauw2018planning}, i.e., how users navigate through urban networks to reach their destinations. Here we focus specifically on the dynamical aspects of mobility and we introduce two general metrics for accessibility of cities: a {\em Velocity Score}, quantifying the overall velocity of access to a specific area of the city, and a {\em Sociality Score} that quantifies how many people one can meet from a specific area.
Finally, the dependence of the \emph{Sociality Score} on the total population of a city can be reduced by scaling it with this quantity. In this way, we define a third accessibility metrics called \emph{Cohesion Score} that quantifies the fraction of the total population that can be met with a typical trip starting from the considered location. The metrics adopted are defined ``general" in the sense that they can be applied in every city and different context allowing comparison between different areas and means of transportation. The proposed metrics allow for an extensive study of the level of accessibility of urban areas, a concept formulated several decades ago and used in different contexts in the literature ~\cite{hansen1959accessibility,black1977accessibility,paez2012measuring, bok2016comparable, boussauw2018planning} to quantify the performance of transportation systems in relations to various aspects of individuals' lives.

There is not just a single definition of accessibility.  Depending on the context, the term accessibility could refer to the availability of services for disabled or disadvantaged people~\cite{schmocker2008mode}, the capability of reaching workplaces for ordinary citizens~\cite{black1977accessibility}, and the possibility of attending certain activities at given times during the day~\cite{miller1999measuring}. Similarly, accessibility can be focused on traveling times using all or several modes of public or private transport or can rely on the spatial distribution of commodities and venues~\cite{geurs2004accessibility}. This proliferation of definitions can make it difficult to reach a unifying view about cities and their dynamical aspects, contributing instead to a dispersion of scientific efforts in diverging directions. The lack of a comprehensive and easy-to-understand definition of accessibility could prevent policymakers from using it in an operational way and scholars from comparing different approaches and methodologies \cite{geurs2004accessibility}.  Our aim here is s to contribute towards a unified and reproducible point of view. 
Thanks to state-of-art routing algorithms, our metrics are designed to be efficiently computed in relatively short times (less than one minute for medium-sized cities). This opens the possibility to explore different scenarios close to real-time. Also, our metrics are well suited for being shared and easily visualized on maps, making them easy to be applied by other researchers to reproduce and extend our results.

The quantification of inequality in accessibility has been proven to be an important tool to assess economic and social inequalities at an extra-urban scale~\cite{weiss2018global}. It is worth mentioning that the local nature of our metrics allows us to evaluate and visualize the geographical fluctuations of the velocity and sociality scores, and thus to quantify the inequalities \indaco{distribution of these measures among areas and population} within each city. In particular, we show that while the distributions of the accessibility metrics seem to have higher values for high-density areas, only a small fraction of the population \indaco{lives in areas with accessibility scores much larger than the rest of the city.} Moreover, the performances of public transport systems decrease in an exponential-like way for all the observed cities with the temporal distance from the city center.
These results exhibit strongly similar patterns among all the observed cities,suggesting the existence of similar causes behind the emergence of this phenomenon, that could range from morphological to Socio-Economic ones.

Despite the local character of the proposed metrics, their aggregation at an urban scale allows for a quantification of the global level of the performances of public services of a city. In this way, the aggregated Velocity Score, the ``City Velocity", represents the overall velocity allowed by the public transportation services. On the other hand, the aggregated Sociality Score, the ``City Sociality", quantifies the number of people possibly met in a standard trip in a given city. The aggregated Cohesion Score, the``City Cohesion",  roughly indicates how well connected a random pair of individuals are in a given city. We adopt these aggregations to rank cities according to public transport performance. We find that, while in general there are correlations between the positions of a city in the different rankings, there are also interesting fluctuations due to the complex interplay between public transport and the population density.

The outline of the paper is as follows. In the Methods section, we illustrate the main tools we adopt throughout the paper, specifically the notion of isochronic map.
We review its definition, and we describe how it is adopted in this paper, including the data and the algorithms to compute it. Based on the computation of these maps, we introduce several accessibility metrics to quantify the efficiency of the public transportation systems and the opportunities provided to the citizens in terms of mobility. The Results section describes several synthetic scores to allow a ranking of cities according to their accessibility patterns. Besides an overall evaluation, we focus in particular on the inequalities of accessibility in cities with respect to their space-time distribution. Finally, we draw some conclusions and highlight interesting future directions.

\section*{Methods}

\label{sec:Accessibility Metrics}

\subsection*{Isochronic maps}

\begin{figure}[h!]
	\includegraphics[width=\textwidth]{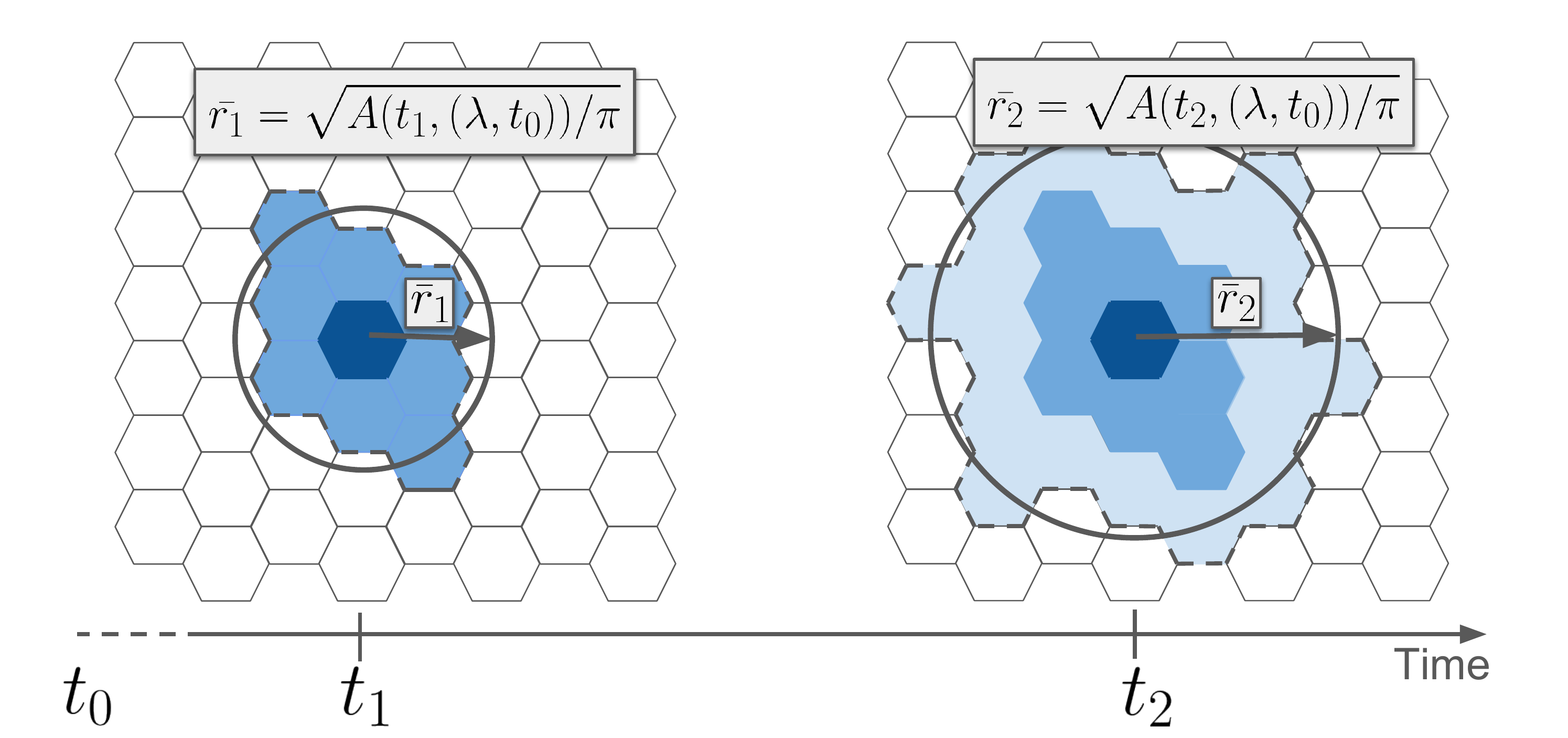}
	\caption{{\bf Isochrone area.} Isochrones with hexagonal tessellation at different times. The circles in figure have the same area of the area contained by the isochrones. }
	\label{fig:circular_iso}
\end{figure}

The accessibility metrics proposed in this paper rely on the notion of \emph{isochronic maps}. An isochronic map is composed by a set of isochrones centered in a given location $\lambda$. The isochrone $I(\tau,\lambda)$ is the contour of the area reachable from $\lambda$ in at most a time $\tau$ and the ensemble of the isochrones obtained for different values of $\tau$ compose the isochronic map of the location $\lambda$.  A more complete definition includes not only the travel-time $\tau$ but also the absolute starting time of the trip. In this way, one has $I(\tau, (\lambda, t_0))$ as the contour of the area reachable from $\lambda$ in at most a time $\tau$ starting at time $t_0$.  Though the notion of isochrones is explored at a quantitative level for a longtime~\cite{Forbes_1964}, it is possible nowadays to compute them massively and very efficiently, opening the possibility for insightful study. The computation of isochrones is based on the computation of the traveling times between any pair of locations in a city using a multi-modal approach that integrates the adoption of all the available public transportation means alternated with walking paths. 
%
In order to keep the computational times low,  we adopted a hexagonal tessellation of the city area which still allows for an exhaustive representation of the public transportation services. We constructed a hexagonal grid with a side of hexagons of $0,2\,$km.  It is worth noticing that not the whole area of a city is covered by hexagons\footnote{We remark that a satisfactory definition of city and its extension is still lacking~\cite{batty2013new}}. We cover with hexagons all locations of a city containing at least a stop of the public service and all areas reachable from any stop of the public service with walking paths not longer than $15\,$minutes.
In order to compute the walking paths between stops of the public service and the hexagonal grid, we use the back-end version of the  Open Source Routing Machine (OSRM)~\cite{luxen-vetter-2011}. The OSRM allows for the computation of shortest walking paths on the urban networks of each city, using the corresponding OpenStreetMap~\cite{OpenStreetMap} network.
As for the schedules of public transit, we relied on data released by public transports companies. Google adopted the GTFS standard file (\url{https://developers.google.com/transit/gtfs/}) to encourage public transport companies to release their data in a uniform way in order to be included in its map platform. It is nowadays possible to find hundreds of companies having released their data, and there are portals where this data is collected and exposed\cite{gtfs_provider}. The databases of public transportation systems are strongly heterogeneous across cities. In some cases, some transportation means could be missing while other extra-urban ones could be included. For instance, for Berlin and London, the GTFS (General Transit Format System) data include all regional trains~\cite{gtfs_provider}. To use a unique and general criterion about the inclusion of areas and transportation means, we adopted the OECD/EU definition of urban areas as {\em functional economic units}~\cite{OECDCities}. The OECD/EU definition exploits the population density to identify an urban core (city core) and travel-to-work flows to identify the hinterland whose labor market is highly integrated with the core (commuting zones). With this definition in mind, we filtered out all the services lying outside both the cores and the hinterland regions from our tessellations. In addition to the database of public transportation systems, we used the population density data on coarse-grained to squares with a surface of $1\,$km$^2$. In order to match the smaller size of the hexagons ($\sim 0,1$km$^2 $) with the size of the square for the population density, we divided the population of each square among the overlapping hexagons proportionally to the fraction of overlapping surface. Data about population densities in urban areas have been gathered through the Eurostat Population Grid~\cite{popGridEurostat} for the European cities~\cite{Eurostat_2017} and the Gridded Population of the world made by the Center for International Earth Science Information Network~\cite{sedacV4}.

The final step to compute the isochronic maps is to put together the coarse-grained representation of a city with the schedule of its public transportation system and to compute traveling times between any pair of hexagons of the tessellation at different times of the day and/or different days of the week. The need for fast commercial transit services has fostered the development of many routing algorithms, capable of computing the optimal routes in urban environments and integrating many different transportation means. Many of these algorithms can perform ``multi-criteria'' optimization, i.e., they can compute the optimal routes minimizing traveling times but also the number of vehicle changes or putting constraints on the arrival times~\cite{delling2009engineering,disser2008multi}. For our purposes, we adopted a modified version of the \emph{Connection Scan Algorithm} (CSA) \cite{dibbelt2013intriguingly}, that we call the Intransitive Connection Scan Algorithm (ICSA). The exact formulation of the algorithm is described in the Appendix.
At a basic level (i.e. not considering walking paths), the CSA features a computation time that scales linearly with the number of connections, i.e. displacements between any two stops of the scheduled public service. The CSA algorithm imposes substantial limitations on the walking path to move from one stop to another. These limitations do not allow its use in a real scenario in urban contexts. Our generalization of the CSA algorithm overcomes these limitations and considers walking paths of less than $15\,$minutes when moving from one stop to another of the public service. Thanks to the ICSA algorithm, it is possible to compute all the shortest-time-paths connecting the centers of any pair of hexagons in the tessellation at several starting times for a typical day of the week. For a typical city with $\approx 10^4$ hexagons, one needs to compute $\approx 10^{8}$ shortest paths. Each one of these shortest-time-paths will consider all the possible means of transport between two hexagons, including the possibility to move on foot to nearby hexagons to access the public transport service places within a given area. The corresponding computational times range between less than two minutes for a medium-size city (like Rome) and about 30 minutes for a big city (for instance New York) on a single CPU of a standard personal computer. The algorithm is easily parallelizable, and it allows to use of the framework described here to implement planning tools where accessibility metrics can be computed in nearly real time (less than one minute of computation). A Python implementation of the computation framework used is released open-source on Github~\url{https://github.com/CityChrone/public-transport-analysis}.

\subsection*{Accessibility Metrics}

In this section, we introduce two universal scores of accessibility that allow for an easy comparison of different areas of the same city and different cities considered as wholes. Interactive representations of those metrics for a large number of cities are available at \url{citychrone.org}.
\indaco{The accessibility quantities proposed aim to measure the performance of public transports at connecting places (velocity score) and people (sociality score). Roughly speaking, the velocity score measures how fast it is possible to reach any point from any other point in the city. The sociality score measures the amount of population that it is possible to reach from any point in the city. Usually, the flow of people in urban systems is described by an origin-destination matrix (ODM). The velocity score can be thought as an accessibility measure that assumes a uniform ODM. Conversely, in the case of the sociality score, we assume an ODM proportional to the population.
}
\begin{figure}[!htbp]
	\includegraphics[width=\textwidth]{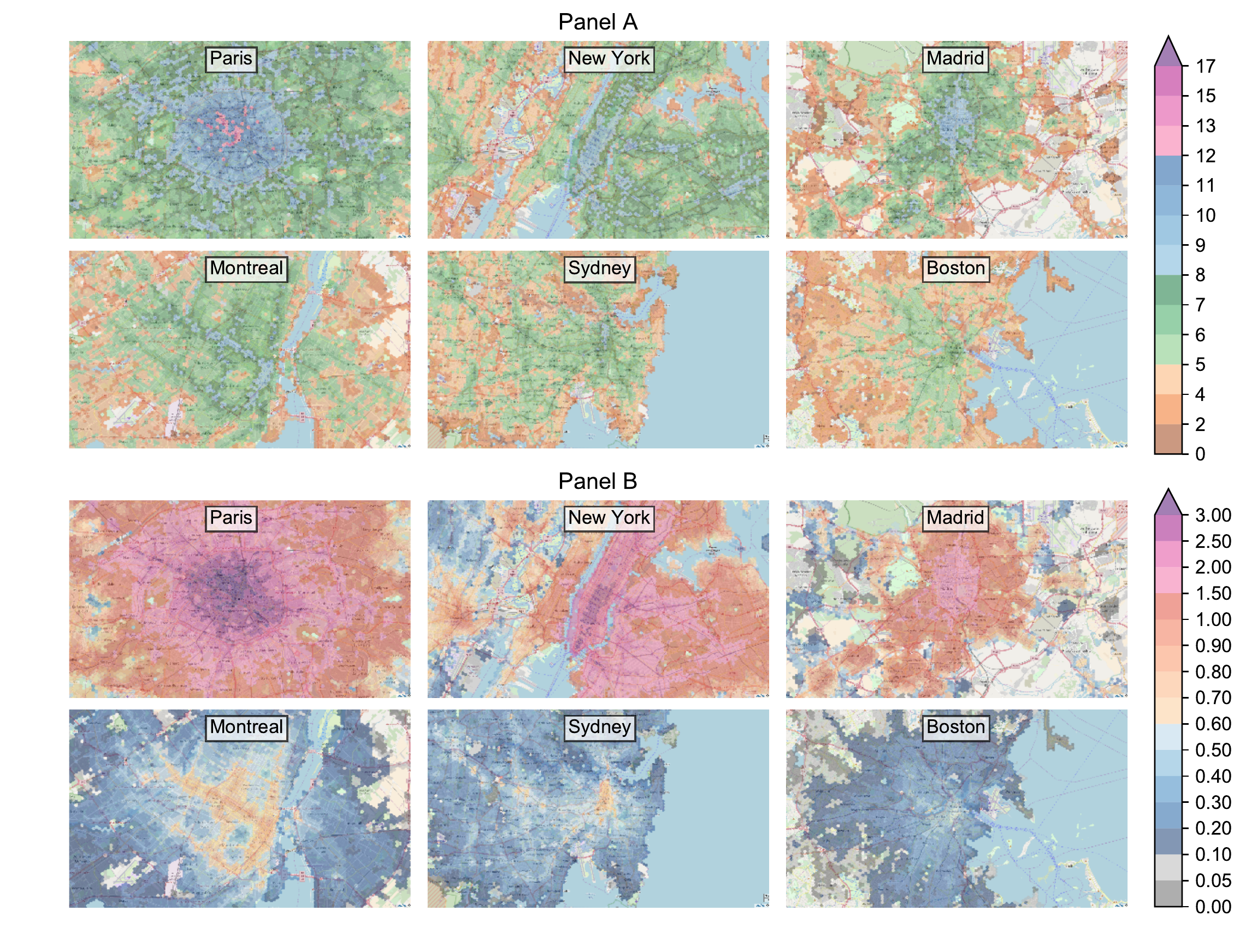}
	\caption{{\bf Maps of the \emph{velocity score} and the \emph{sociality score}} In the six maps of panel A (B) we report the \emph{velocity score} in km/h (\emph{sociality score} in millions of inhabitants) for six different cities: Paris, New York, Madrid, Montreal, Sidney, Boston. The values of the \emph{velocity score} range from less than $2\,$km/h (brown) of velocity score up to more than $17\,$km/h (purple) whereas ranges from less $0.05$ millions of inhabitants reachable up to more than $3$ millions of individuals for the sociality score. The great variability of the colors reveals a strong dissimilarity of performances of the public transports across cities.}
	\label{fig:velocity_score_cities}
\end{figure}

\subsubsection*{Velocity score} 
\label{sec:velocity_score}

The {\em velocity score} aims at giving a synthetic representation of the information encoded in all the isochronic maps computed from all the points of a city. To this end, we imagine the isochronic map as a spreading process from a starting point, and we are interested in the average speed of expansion of the front of the isochrone as a function of time.
More precisely, let us consider the isochrone centered at the hexagon $\lambda$ at time $t_0$ corresponding to a travel-time $\tau$, $I(\tau, (\lambda,t_0))$. The \emph{covered area} $A(\tau, (\lambda, t_0))$ of the isochrone at time $\tau$ will thus be the area contained within $I(\tau, (\lambda,t_0))$. By approximating the perimeter of the isochrone with a circle, the average traveled distance $\bar{r}$ taking a random direction from the starting point $p_0$ is given by:

\begin{equation}
	\bar{r}(\tau, (\lambda, t_0)) = \sqrt{A(\tau, (\lambda, t_0)) / \pi },
\end{equation}

\noindent and dividing by the time $\tau$ we obtain a quantity that has the dimension of a speed:

\begin{equation}
	\bar{v}(\tau, (\lambda, t_0)) = \bar{r}(\tau, (\lambda, t_0))/\tau.
\end{equation}

\noindent The interpretation of $\bar{v}(\tau, (\lambda, t_0))$ is the average speed of expansion, at time  $\tau$, of a circular isochrone with the same area of the real one (see Fig.\ref{fig:circular_iso}).

This quantity can be considered approximately as the average velocity of a journey of duration $\tau$ choosing a random direction from the starting point. On the other hand, this quantity is proportional to the square root of the amount of area it is possible to explore from the hexagon $\lambda$ given a time interval of $\tau$. We chose to consider the square root of the area instead of the area itself to have a more direct interpretation of it in terms of transportation velocity, because it is easier to communicate and to understand for a general audience. This quantity is defined for every hexagon $\lambda$ and any starting time $t_0$ and travel-time $\tau$. The \emph{velocity score} is obtained by averaging over both the starting time $t_0$ and the travel-time $\tau$, as: 

\begin{equation}
	\label{eq:velocity_score}
	v(\lambda) = \frac{\sum_{t_0 = 6am}^{10pm}  
	\int_0^{\infty} v(\tau, (\lambda, t_0)) 
	f(\tau) d\tau}{\sum_{t_0 = 6am}^{10pm} 
	\int_0^{\infty} f(\tau) d\tau},
	\label{eq:vel_score_avg}
\end{equation}

\noindent where several starting times have been considered, from 6 am to 10 pm with a step of 2h. In equation (\ref{eq:vel_score_avg}) the average over $\tau$ is performed by weighting with a travel-time distribution $f(\tau)$. The travel time distribution represents the probability for an individual or a group of individuals to perform a journey of duration of $\tau$. The travel time distribution could vary between the considered cities, time frames~\cite{geurs2004accessibility}, and also between areas and groups of individuals of the same city~\cite{Accessibility_Houston}. In the Appendix we show how the \emph{Velocity Score} (and the other accessibility metrics defined in the following) computed with different choices for $f(\tau)$ are highly correlated with one another. Thus, the choice of $f(\tau)$ does not alter qualitatively the results obtained. On the other hand, using the same $f(\tau)$ for each city is equivalent to focusing on the perspective of a single individual, or a cohesive group of individuals, who would compare different cities and different transportation systems from their perspective. For all these reasons we focused on one specific travel time distribution, namely that obtained from fits of surveys of the daily budget times spent on a bus by UK citizens~\cite{kolbl2003energy}. We remark that, though out of the scope of the present paper, the investigation of the impact of different city-specific travel time distributions deserves further investigation.
\newline
Fig.\ref{fig:velocity_score_cities} (panel A) shows the velocity scores of six different cities. For interactive explorations of the maps and other cities we refer the reader to the platform \url{citychrone.org}.

\subsubsection*{Sociality score}

The \emph{velocity score} introduced above represents an indicator of how good the public service is at allowing a fast exploration of the urban space. At this stage, this score does not take into account the population density distribution. We know instead that there is a strong interplay and feedback loop between the efficiency of the public service and the population density. While it is normal to strengthen the service in highly populated areas, regions with a low population density risk being poorly served by public transportation. In order to quantify this interplay, we introduce a second metrics that quantifies the performance of public transit in connecting people. 
Let us now define  $P(\tau, (\lambda, t_0))$ as the amount of population living within the isochrone $I(\tau, (\lambda, t_0))$. Similarly to what we did for the velocity score, we can average $P(\tau, (\lambda, t_0))$ over the travel time $\tau$ (with the same distribution of daily budget times $f(\tau)$) and over different starting times $t_0$, obtaining the {\em sociality score} as:

\begin{equation}
	s(\lambda) =  \frac{\sum_{t_0 = 6am}^{10pm} \int_0^{\infty} P(\tau,(\lambda, t_0))f(\tau)d\tau}{\sum_{t_0 = 6am}^{10pm} 
	\int_0^{\infty} f(\tau) d\tau},
	\label{eq:sociality_score}
\end{equation}

\noindent 
Considering a typical working day, the Velocity Score provides an approximate measure of the average speed at which an individual can move away from a hexagon $\lambda$, in a randomly chosen direction. Instead the Sociality Score provides a measure of the number of people it is possible to reach within the same trip. 
\indaco{The sociality score can also be interpreted as a measures of the amount of the population that can easily reach the point considered, assuming that, on average, the travel time of trips in cities is similar reversing origin and destination. In order validate this assumption we compute the sociality score with travel time of the incoming trip and out coming trips for each point in Rome. In fig.5 in the Appendix there is the scatter plot of this two quantities showing the high correlation between these two measures.} Then the fig.~\ref{fig:velocity_score_cities} (Panel B) shows the Sociality Score maps for the same cities considered for the velocity score. 

\section*{Results}

\subsection*{City rankings}

The scores introduced above allows us to rank cities according to the overall performances of their public transport system. 
To this end, we introduce the {\em City Velocity} indicator as the average {\em Velocity Score}, weighted over the population density. The second indicator we introduce is the {\em City Sociality}, defined as the average {\em Sociality Score} weighted over the population density. While the City Velocity is a measure of the how fast a typical inhabitant can visit the city on a typical trip, the City Sociality is a measure of the how many distinct people it is possible to meet. Finally, we introduce the {\em City Cohesion} indicator, which measures the easiness for two randomly picked individuals to meet within a city. The larger this indicator is, the more the city is cohesive and favors social interactions among its citizens. 
We note that the assumption of using the same travel-time distribution $f(\tau)$ for each city is quite strong since citizens of different cities might exhibit different travel habits. However, we are focusing on the perspective of a single individual, or cohesive group of individuals who would compare different city, as explained in the subsection \emph{Accessibility Metrics}. In the Appendix we show how the rankings weakly depend from a reasonable choice of travel time distributions.
\subsubsection*{City velocity}
\begin{figure}[!ht]
	\includegraphics[width=\textwidth]{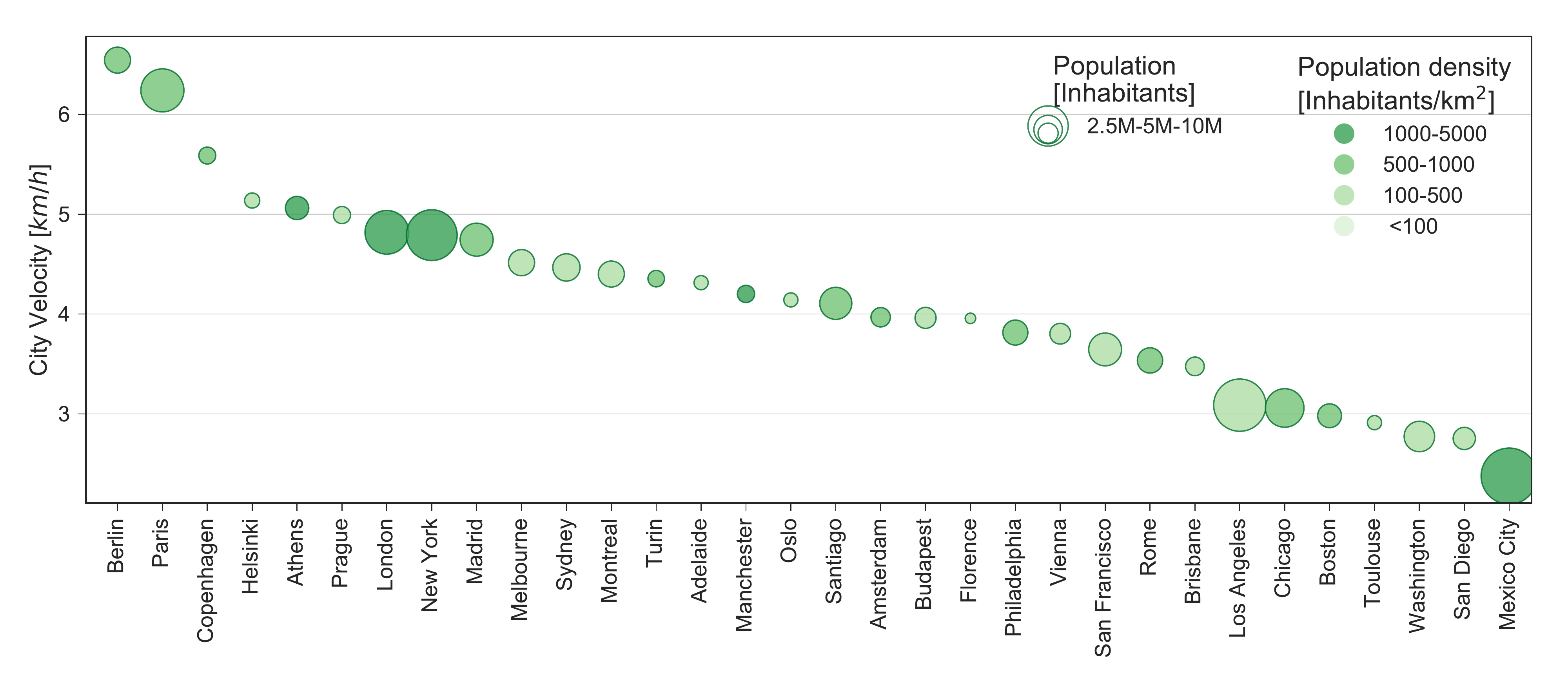}
	\caption{{\bf Ranking of cities according to the City Velocity defined in Eq.~\ref{eq:city_velocity}}. Cities are displayed with circles whose size is proportional to the total population and whose saturation of the filling color is proportional to the overall population density.}
	\label{fig:city_velocity}
\end{figure}

For each hexagon, $\lambda$, we have both the number of people living there, $pop(\lambda)$, as well as the average velocity of their trips with public transports starting from the considered hexagon, $v(\lambda)$ (Eq.~\ref{eq:velocity_score}). In this way, we can compute the average velocity per person of the whole city, representing the average amount of different places a typical person living in the city can easily access with public transit. In particular, we define the {\em City Velocity} as the average velocity per person:

\begin{equation}
v_{city} = \frac{\sum_{\lambda \in city} v(\lambda) * pop(\lambda)} {pop(city)}
\label{eq:city_velocity}
\end{equation}
\noindent where $pop(\lambda)$ is the population in the hexagon $\lambda$. In the equation \ref{eq:city_velocity} we sum over all the hexagons in the city weighted by the population living in that hexagon, and we divide by the total of the population of the city (living in the core and the commuting zones), $pop(city)$.  Notice that we assign zero velocity to all the areas of the city not covered by hexagons, i.e., the areas more than $15$ minutes away from any stop. Fig.\ref{fig:city_velocity} reports the ranking of several cities according to their City velocity.  The highest ranked cities are Berlin and Paris, with values 20\% higher than any other city. This means that typically a citizen of Berlin and Paris can explore the space around at least  $20\%$ faster than the others. Copenhagen, Helsinki, Athens, Prague, London, and New York features good performance. On the other side of the spectrum, Mexico City, San Diego, and other U.S. cities have a large fraction of the population \indaco{with very low velocity score.} 

\subsubsection*{City sociality}

\begin{figure}[ht!]
	\includegraphics[width=\textwidth]{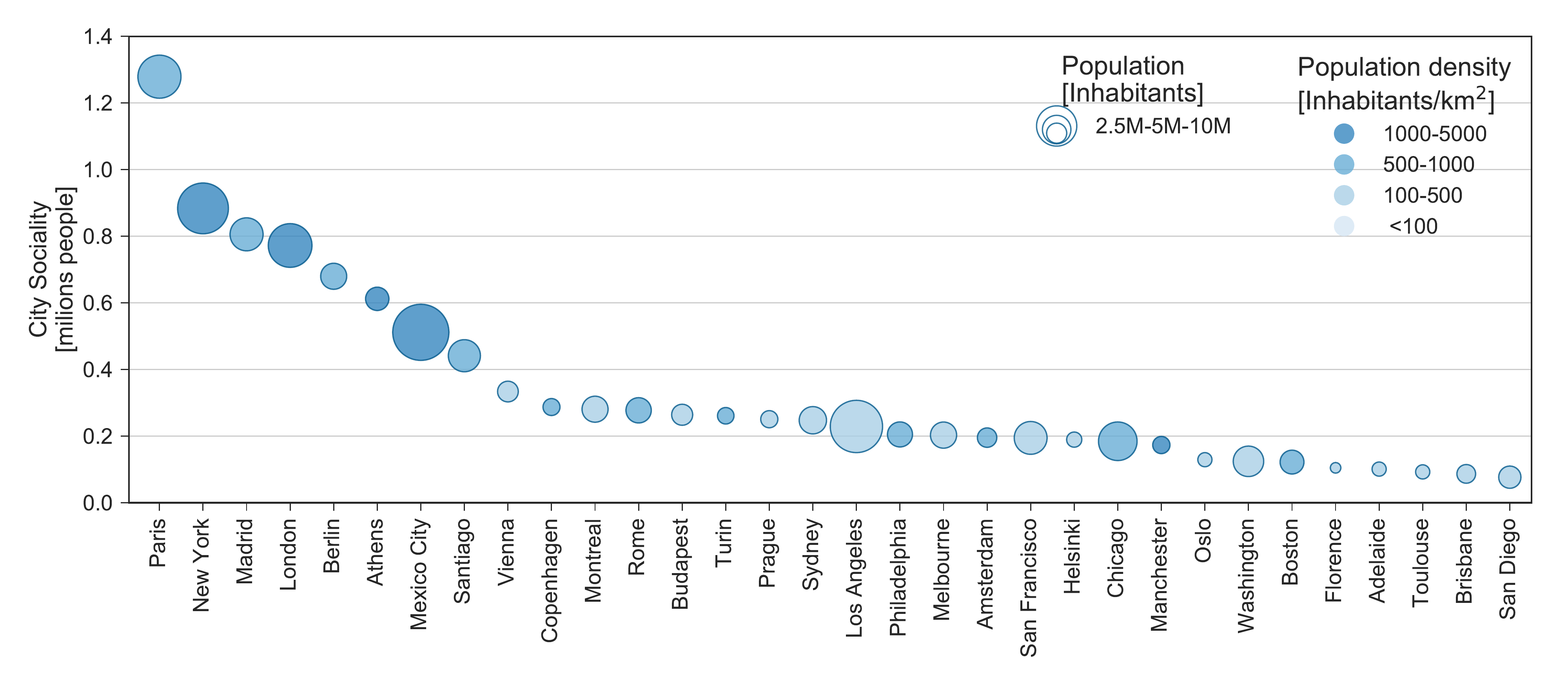}
	\caption{{\bf Ranking of cities according to the City Sociality defined in Eq.~\ref{eq:city_sociality}}. Cities are displayed with circles whose size is proportional to the total population and whose saturation of the filling color is proportional to the overall population density.}
	\label{fig:city_sociality}
\end{figure}

The {\em City Sociality} is defined as:

\begin{equation}
s_{city} = \frac{\sum_{\lambda \in city} s(\lambda) * pop(\lambda)  }{pop(city)}.
\label{eq:city_sociality}
\end{equation}

\noindent  As for the {\em City Velocity}, we average over the population distribution, and the areas of the city not served by public transport are considered to have zero {\em Sociality Score}. The {\em City Sociality} is the typical number of people that a person living in the city can potentially meet within a typical daily trip. The ranking of cities, according to the City Sociality, reported in Fig.\ref{fig:city_sociality}, features some differences for the corresponding ranking obtained with the City Velocity. In this case, Paris gains the first position thanks to its high population density in the city core and its efficient and capillary public transit system. Among the set of considered cities, Paris is the only one where on average a person can potentially meet over one million people in a typical daily trip. Scrolling the ranking, the City Sociality decreases initially quickly, with the most populated cities in the first positions, then eventually decreases very slowly for smaller cities.

\subsubsection*{City cohesion}

By re-scaling the {\em City Sociality} with the total population of a city, we obtain the {\em City Cohesion}:

\begin{equation}
c_{city} = \frac{s_{city}}{pop(city)}.
\label{eq:city_cohesion}
\end{equation}

\begin{figure}[htbp!]
	\includegraphics[width=\textwidth]{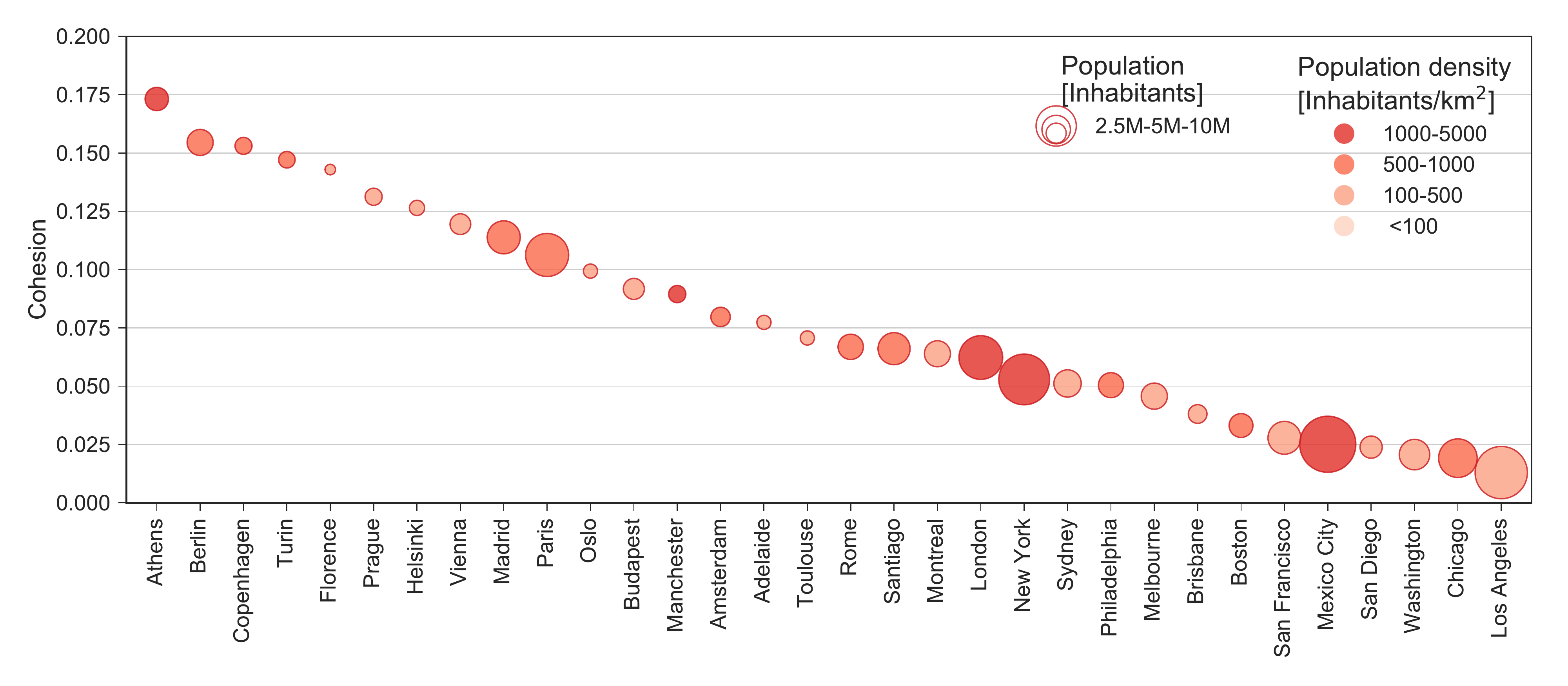}
	\caption{{\bf Ranking of cities according to the City Cohesion defined in Eq.~\ref{eq:city_cohesion}}. Cities are displayed with circles whose size is proportional to the total population and whose saturation of the filling color is proportional to the overall population density.}
	\label{fig:cohesion_score}
\end{figure}

\noindent The {\em City Cohesion} gives an estimate of the fraction of the population that can be reached by a typical trip of an inhabitant of the city. Fig.~\ref{fig:cohesion_score} shows the ranking of cities according to their {\em City Cohesion}. The first city is Athens, thanks to a good public transportation system and a very high-density population concentrated in the core of the city. In second and third positions are Berlin and Copenhagen, which also features very high \emph{Velocity Scores}. Then we find Turin and Florence featuring a right balance between the population distribution and the efficiency of the public transportation system, despite relatively low \emph{City Velocity} and \emph{City Sociality}. A large part of US cities have a low City Cohesion score, resulting from the low population density in the city core, making those cities very dispersive.

\subsection*{Inequalities in urban accessibility patterns}

In this section, we focus on a particular aspect of accessibility, \indaco{the spatial-temporal distribution inside the city.} A high position of a city in the overall ranking for any of the scores presented above does not imply {\em per se} that the same accessibility patterns are granted to all citizens. In order to investigate dis-homogeneities in the accessibility patterns, one needs to take a closer perspective and look at the accessibility metrics at a more fine-grained scale within cities. We focus for visualization clarity reasons on a subset of cities, namely the same cities we focused on in the section devoted to Accessibility Metrics: Paris, New York, Madrid, Montreal, Sydney, Boston. In the Appendix we show how the results presented are valid also for the other city analyzed.

\begin{figure}[ht!]
	\includegraphics[width=\textwidth]{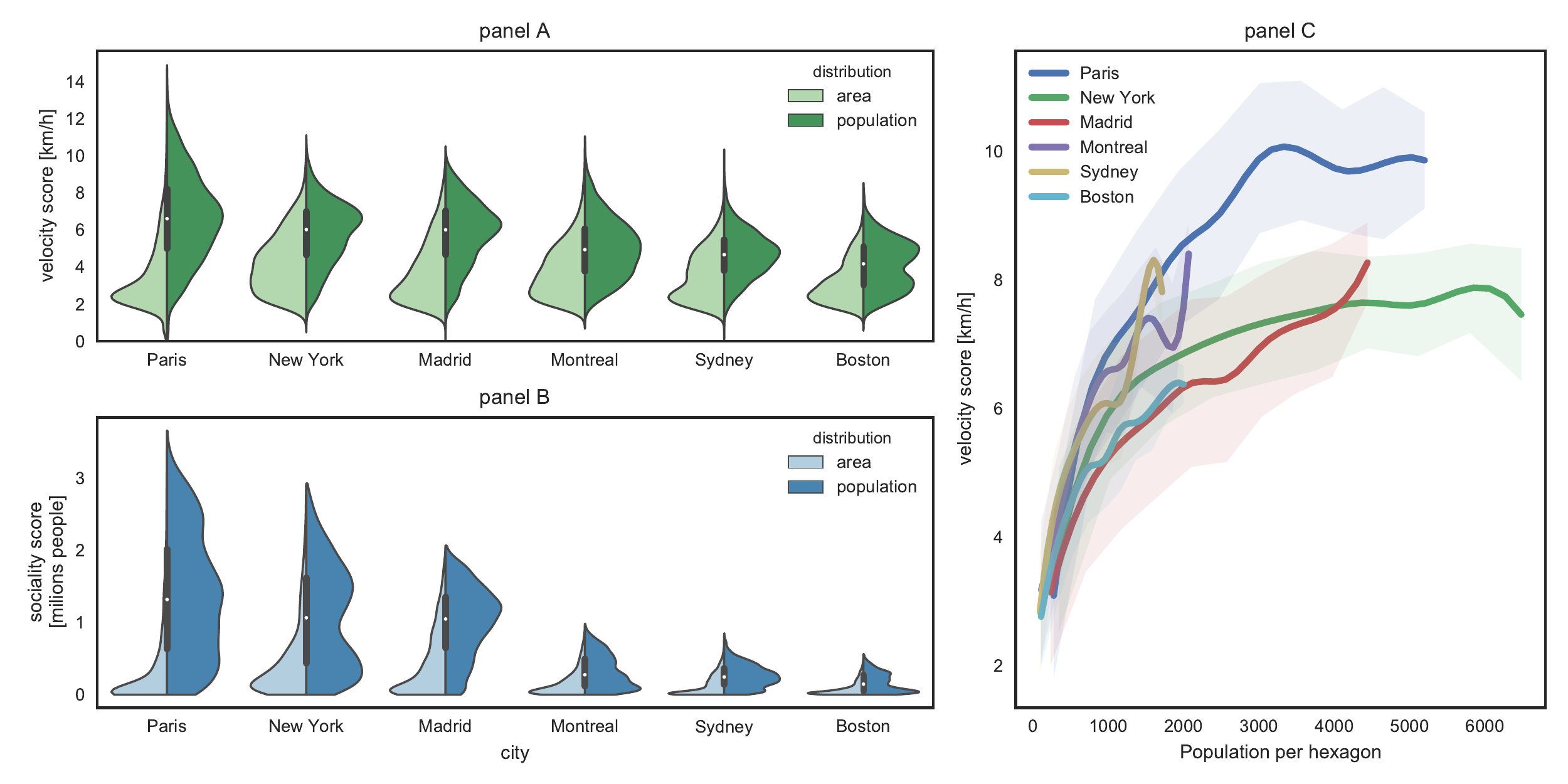}
	\caption{{\bf Distributions of the Velocity (panel A) and the Sociality (panel B) scores and velocity score vs. the population density (panel C)} {\bf Panel A (panel B}): Distribution of the {\em velocity score} ({\em sociality score}). The light green (blue) area represents the distribution of the areas featuring a given value of the {\em velocity score} ({\em sociality score}). The dark green (blue) area represents the distribution of population with a given value of the  {\em velocity score} ({\em sociality score}). The distribution for the population density is peaked towards higher values than that related to the area, signaling the fact that denser areas are associated, on average, to better public transit systems. We do not report here the results for the Cohesion metrics since it would give the same information of the Sociality Score. {\bf Panel C}. The average value of the velocity score in hexagons with a given population for the six cities of Paris, New York, Madrid, Montreal, Sydney, and Boston. In all cases, one observes an increasing trend. The shadows around the average values curve represent the standard deviation of the velocity score distribution for each corresponding value of the population}
	\label{fig:violin_plot}
\end{figure}

An interesting way to represent the Velocity and Sociality score is through a violin plot, as reported in Fig.~\ref{fig:violin_plot}.

Panels A and B refer to the distributions of the Velocity and Sociality scores, respectively. The way in which one reads these plots is the following. For each city we plot the distribution of areas and population as a function of the Velocity or Sociality score. For instance, panel A refers to the Velocity score. For each city, we plot in light green the normalized distribution of areas (hexagons) as a function of the Velocity score, i.e., the fraction of hexagons featuring a specific value of the Velocity score. A very efficient city has this distribution peaked around high values of the Velocity score. From this perspective, New York appears to have the most balanced distribution of Velocity scores across its whole area. 
On the other hand, represented in dark green is the distribution of the population density as a function of the Velocity score, i.e., the fraction of the population associated with a specific value of the Velocity score. A city, with well-distributed public transport accessibility among the population, has this distribution peaked around high values of the Velocity score. From this perspective, Paris, New York, and Madrid appear to have more equally distributed velocity scores than Montreal, Sydney, and Boston. Panel B reports the same information as panel A (in light and dark blue) for the Sociality score. The difference between Paris, New York, and Madrid, on the one hand, and Montreal, Sydney, and Boston, on the other, in terms of the range of sociality score both for areas and population is striking. Paris and New York appear to feature the broadest distribution of Sociality scores across their citizens. 

It is evident, both in panels A and in B, that  (light green and blue areas) a large number of hexagons within the city borders display low values of the accessibility scores. However, when the population density is taken into account (dark green and blue areas), the peaks of the distribution shift towards high-populated areas. This result is somehow unsurprising, considering that the public transportation systems are mainly designed to serve the largest amount of citizen as possible as allowed by the limited financial resources.  Fig.~\ref{fig:violin_plot} (panel C)  confirms this picture, where it is evident the growing trend of the average values of the Velocity scores at fixed population density with the population density for the six cities considered above.

\begin{figure}[ht!]
	\includegraphics[width=\textwidth]{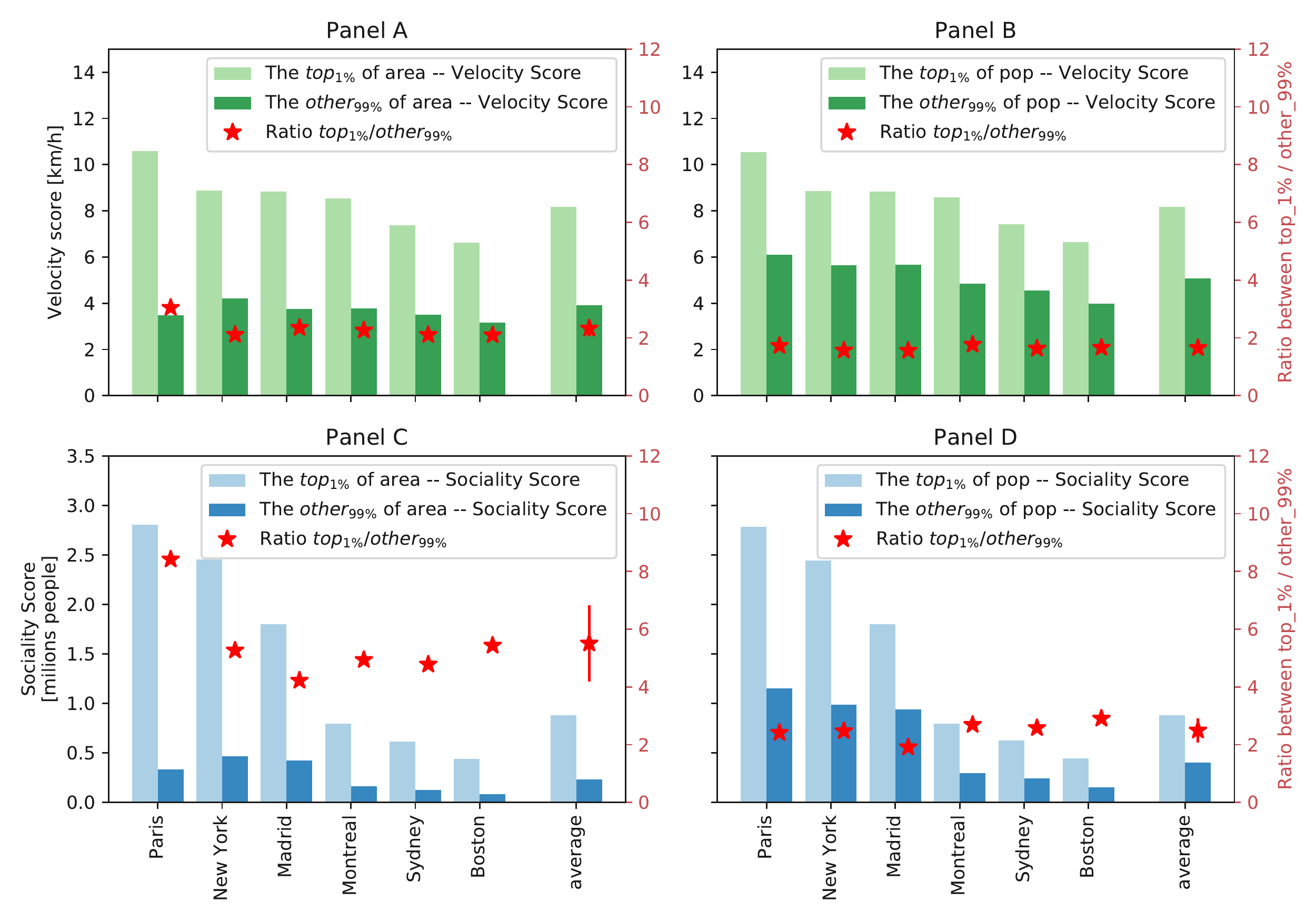}
	\caption{{\bf Inequalities in accessibility patterns.} {\bf Panel A}: Average values of the velocity scores among the hexagons featuring the top $1\%$ (light green) of the values of the velocity score as compared to the remaining $99\%$ (dark green) for the six selected cities. The last columns report the same values averaged over all the 32 analyzed cities. Red star marks on the right $y$-axis are the ratios between the average values of top $1\%$ and the other $99\%$. {\bf Panel B}: Average values of the velocity scores among the population with the highest top $1\%$ (light green) of the values of the velocity score as compared to the remaining $99\%$ (dark green) for the six selected cities. The last columns report the same values averaged over all the 32 analyzed cities. Red star marks on the right $y$-axis are the ratios between the average values of the top $1\%$ and the other $99\%$. {\bf Panel C}: Same as Panel A for the Sociality score. {\bf Panel D}: Same as Panel B for the Sociality score.}
	\label{fig:one_ninenine}
\end{figure}

The trends reported above, correlating denser populated areas to higher (on average) accessibility scores, do not imply that the planning of public transportation systems succeeds in reducing inequalities in the accessibility patterns. The spread of the distributions is still very high and very few people (or areas) have access to high \indaco{ accessibility values} compared to the rest of the populations (or areas). This is true for all the accessibility scores introduced above. In order to better quantify the large variability of urban accessibility patterns, we divide urban areas (hexagons) and population in two classes: hexagons and people featuring the top $1\%$ of values of the Velocity and the Sociality scores and the remaining $99\%$. For each of the two classes, we compute the average values of the Velocity and the Sociality scores and we compare them. The results are reported in Fig.~\ref{fig:one_ninenine}: panels A and B for the Velocity score and panels C and D for the Sociality score, panels A and C for the distribution of hexagons and panels B and D for the population densities. 
The striking though perhaps not surprising result confirms the strong level of inequalities observed for all the cities considered. The ratio between the average values of the scores of the two classes is always larger than two. This implies that focusing for instance on the velocity score, the top $1\%$ of the hexagons (populations) features values of the velocity score that are double the remaining $99\%$.  
In other words, $1\%$ of the city areas allow for daily trips at twice the speed of the rest of the city, and $1\%$ of the population can move around at least twice as fast as the rest of the population. 
Similar considerations hold for the sociality scores, which implies that $1\%$ of the population potentially has access at twice the number of \indaco{people} as the rest of the population. The ratio between the values of the top $1\%$ compared to the remaining $99\%$ is similar across all considered cities (see Appendix, Figure 5 and Figure 6), as witnessed by the error bars reported in Fig.~\ref{fig:one_ninenine}. It is also interesting to observe that almost all the ratios between values of the average scores computed for the two classes ($1\%$ and $99\%$) lie between $2$ and $4$, suggesting the existence of general patterns of organization across very different cities and urban environments.

\subsubsection*{Space-time distribution of inequality in accessibility patterns}

\begin{figure}[t]
	\includegraphics[width=\textwidth]{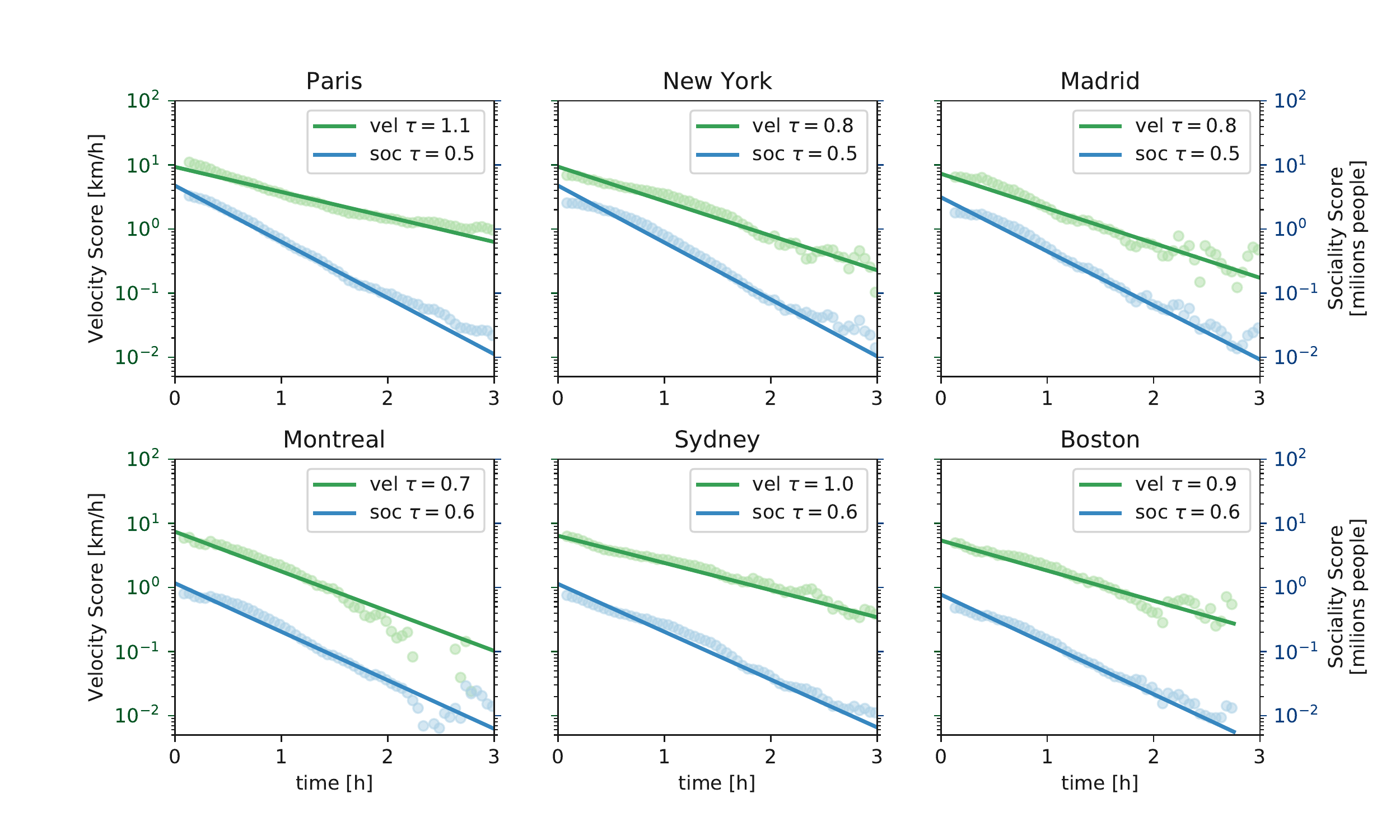}
	\caption{{\bf Exponential decay of the velocity score and the sociality score with travel-times from the city center.} Average values of velocity scores (green points) and sociality scores (blue points) at a given travel-time distance from the hexagon with the highest score in each of the six selected city. The lines are the best fit of the data with the function (\ref{eq:fit_eq}). In the legend for each figure, the parameters $\tau$ of equation (\ref{eq:fit_eq}) found for the decay of velocity score and the sociality score are reported. }
	\label{fig:exp_vel}
\end{figure}

The quantitative assessment of the strong \indaco{uneven distribution} observed in the accessibility patterns reported above can be further clarified by looking at the spatial distribution of the accessibility metrics. In the maps shown in Fig.~\ref{fig:velocity_score_cities} (and at www.citychrone.org), we observe a central area with the highest values of the accessibility observables and some ``islands'' with high accessibility values connected to the central zone by some well-served directions, consistent with the idea of polycentric cities\cite{clark2000monocentric}. To better quantify this effect, we show the behavior of the velocity and the sociality scores (Fig.~\ref{fig:exp_vel}) as a function of the travel time from the center of each city. Here the center of a city is defined as the hexagon with the highest score (velocity and sociality, respectively).

Both the velocity and the sociality scores decay fast as a function of the travel time from the city center. The exponential function well describes this decay:
\begin{equation}
f(t) = \sigma_{0} e^{-\frac{t}{\tau}} + \sigma_{\infty},
\label{eq:fit_eq}
\end{equation}    

\noindent where $\tau$ represents the typical decay time, $\sigma_{\infty}$ is the lower bound of the velocity score for each city, and $\sigma_{0}$ represents the average velocity score of areas (hexagons) nearby the best performing one. We performed the best fit by binning $\tau$ in order to remove biases coming from better sampled temporal distances.
\noindent The value $\sigma_{\infty}$ represents the value of the score (either velocity or sociality) acquired in hexagons at the temporal edge of the city itself, i.e., for the farthest (in travel-time) hexagons from the city center. Hence we estimated it as the average velocity score (sociality score) of the $5\%$ least accessible hexagons from the city center. The parameters $\tau$ and $\sigma_0$ are obtained through a linear regression of the quantity $\log (f(t) - \sigma_{\infty})$, which depends linearly on $t$. The curve well fits the decay of the mean values of the velocity score, the average value among all $32$ cities analyzed of R-square is $R^2 = 0.92$ (see Appendix, Figures 7 and 8). The average value of the characteristic time $\tau$ is $0.86\;hours$, ranging from $0.4\;$h for Santiago to the $1.6\;$h for Los Angeles. The dependence of the sociality score on the temporal distance from the best performing hexagon of the city is again well-described by the equation (\ref{eq:fit_eq}), with a value of R-square $R^2 = 0.95$ higher with respect to those of the velocity score and an average value of $\tau=0.55\;$h. The smaller characteristic time is due to the convolution of the decay of the velocity score with the well-known decay of the population density from the city center, which is again exponential\cite{clark_1951,muth_1969,mills_1970}.
The ensemble of the results confirms that inequalities in the accessibility patterns allowed by public transport favor a small portion of the total city area and a small fraction of the population, typically clustered around certain areas. Moving in space and time away from these areas will lead to experiencing generally \indaco{much less performing } public transport services. This behavior and the stability of inequality patterns of Fig.~(\ref{fig:one_ninenine}) strengthen the hypothesis of shared causes, independent from the particular location, behind the emergence of the decay of public transport performances that will deserve future investigations.
\section*{Discussion}
The study of accessibility in urban contexts represents a multifaceted topic whose relevance transcends the mere problem of optimizing transportation systems, though this a very complex. It impacts the level of opportunities available in a city, the equal access to them, the level of inclusion of minorities. The interpretations that one can give to the notion of accessibility can take are numerous: from the planning of better and more efficient urban environments to the improvement of quality of life in rural areas, from the definition of real estate market prices to the definition of new business models of mobility and so on. The extreme generality of the term accessibility also depends on the specific aspects one could be interested in: the availability of jobs in a specific area, the quality of the schools in a neighborhood, the possibility to take part in leisure activities depending on the time of the day. Despite a long tradition of scientific studies on these subjects, no consensus has yet emerged on how to quantify accessibility in a general way, i.e., through metrics applicable in very many situations and very different urban contexts. 
\indaco{The main aim of this paper is to give a contribution towards a unifying, simple and general framework for accessibility studies. We proposed some general metrics that allow for a quantitative comparison of different cities and different areas of the same city. Despite the limitations of some of our assumptions, our framework and measures are easily reproducible and applicable to the study of accessibility via public transport in every urban environment in which transit feed open data is available.}
To this end, we took a specific angle by looking at the city and at the paths within it from the point of view of traveling-times, which allows mapping the city in a way much closer to individuals' perception. The cornerstone of this approach is the computation of isochronic maps and, based on them, the introduction of several scores that take into account the performance of public transport to connect areas and people. The primary outcome is a set of scores that quantify how well served is a city from the public transit and how well a specific area of the city is connected to the rest of the city. We show how these scores allow comparing the performances of public transport systems of different cities around the world, pointing out the differences in their ability to expand the range of opportunities and enhance social interactions. A very interesting opportunity open by the new scores concerns the possibility to quantify the level of \indaco{uneven distribution of these quantities within a city, i.e., the fluctuations of the accessibility scores among areas and population.  We remark here that our first aim was to measure the performance of public transport to connect places and people. Despite that more realist origin destination matrix, for instance considering the distribution of opportunities within the city, can be considered and easily integrated into our framework. But, up to now, there is still a lack of open data-sets covering enough cities for this kind of analysis.}  
\indaco{Taking into account the aim of the accessibility measure proposed, our} analyses reveal a general pattern observed in all the considered cities. Namely that the $1\%$ of the area of a city features accessibility scores with average values at least double those of the remaining $99\%$ of areas. 
The same patterns are observed by looking at the number of people enjoying specific values of the accessibility scores: also in this case, the top $1\%$ of the population can move at least twice as fast as the remaining $99\%$ of the population. This very uneven distribution of performances of the public transport within an urban environment is explained in terms of the rapid decay of the accessibility scores as a function of the temporal distance from the city center. The observed similarities of the mobility patterns across different cities suggest the existence of common causes, independently of the specific location. The observed inequality patterns are the results of the planning and organization of public transport systems. \indaco{ Considering our initial remarks, we can speculate that these patterns might be explained by the limited resource urban planners have to deal with when designing public services. In this sense, including important locations and fluxes might allow us to understand if these resources are efficiently allocated to satisfy the mobility needs of the citizens.}
The availability of general scores for accessibility and inequalities could be the first step towards a more systematic evaluation of the present situation in urban contexts and careful planning of future scenarios. The \url{citychrone.org} platform is a relevant example in that direction, already allowing for both the visualization of all the accessibility metrics introduced here and the conception of new scenarios for improved mobility and accessibility. \indaco{As a final remark, the inclusion of other data sources, such as  points of interest in the accessibility metrics (e.g. workplaces, shops, schools, etc.) or considering fluxes of people is quite straightforward in our framework and could lead to interesting results either in the global ranking of accessibility between cities and in the comparisons between city areas, by giving more importance to the purpose and popularity of certain trips.}

\section*{Data, code and materials}

All the data used in this work can be freely accessed from public repositories \cite{OECDCities, gtfs_provider,popGridEurostat,sedacV4}. Python source code used to compute the accessibility quantities and for the analysis performed are freely downloadable from the online \emph{public-transport-analysis} GitHub repository [DOI:10.5281/zenodo.1309835]\cite{db1}. The hexagons tessellation and the related accessibility quantities computed can be download from the online  \emph{openData} GitHub repository [DOI: 10.5281/zenodo.1309927]\cite{db2}. In the same repository, there is also a CSV file (agency.csv) with the list of public transports agency used for each city.

\section*{Funding}

The authors acknowledge support from the project KREYON, funded by the John Templeton Foundation under contract n. 51663.  The authors acknowledge Sapienza University of Rome and ISI Foundation for the support in the management of the funding. This work has been partially  supported by the SmartData@PoliTO center on Big Data and Data Science, and by Sony Computer Science Labs Paris.

\section*{Competing Interests}

The authors declare no competing interests.

\section*{Author contributions statement}

I.B., V.L., and B.M. designed the research; I.B. performed the analysis; I.B., V.L., and B.M. analysed the results and wrote the paper.


\newpage

\begin{thebibliography}{10}
	
	\bibitem{nations2014world}
	of~Economic D, Social~Affairs PD.
	\newblock {World Urbanization Prospects: The 2014 Revision, Highlights
		(ST/ESA/SER.A/352)}.
	\newblock Population Division, United Nations. 2014;Available from:
	\url{http://www.un.org/en/development/desa/news/population/world-urbanization-prospects-2014.html}.
	
	\bibitem{batty2013new}
	Batty M.
	\newblock {The new science of cities}.
	\newblock Mit Press; 2013.
	
	\bibitem{mayer2013big}
	Mayer-Sch{\"o}nberger V, Cukier K.
	\newblock {Big data: A revolution that will transform how we live, work, and
		think}.
	\newblock Houghton Mifflin Harcourt; 2013.
	
	\bibitem{Gallotti2016}
	Gallotti R, Bazzani A, Rambaldi S, Barthelemy M.
	\newblock {A stochastic model of randomly accelerated walkers for human
		mobility}.
	\newblock Nature Communications. 2016 08;7:12600.
	\newblock Available from: \url{http://dx.doi.org/10.1038/ncomms12600}.
	
	\bibitem{Alessandretti160156}
	Alessandretti L, Karsai M, Gauvin L.
	\newblock User-based representation of time-resolved multimodal public
	transportation networks.
	\newblock Royal Society Open Science. 2016;3(7).
	\newblock Available from:
	\url{http://rsos.royalsocietypublishing.org/content/3/7/160156}.
	doi:10.1098/rsos.160156.
	
	\bibitem{1367-2630-5-1-348}
	K\"olbl R, Helbing D.
	\newblock {Energy laws in human travel behaviour}.
	\newblock New Journal of Physics. 2003;5(1):48.
	\newblock Available from: \url{http://stacks.iop.org/1367-2630/5/i=1/a=348}.
	
	\bibitem{mastroianni2015local}
	Mastroianni P, Monechi B, Liberto C, Valenti G, Servedio VD, Loreto V.
	\newblock {Local Optimization Strategies in Urban Vehicular Mobility}.
	\newblock PloS one. 2015;10(12):e0143799.
	
	\bibitem{fleurquin2014characterization}
	Fleurquin P, Ramasco JJ, Egu{\'\i}luz VM.
	\newblock {Characterization of delay propagation in the US air-transportation
		network}.
	\newblock Transportation journal. 2014;53(3):330--344.
	
	\bibitem{Gallotti2015}
	Gallotti R, Bazzani A, Rambaldi S.
	\newblock {Understanding the variability of daily travel-time expenditures
		using GPS trajectory data}.
	\newblock EPJ Data Science. 2015;4(1):18.
	\newblock Available from:
	\url{http://dx.doi.org/10.1140/epjds/s13688-015-0055-z}.
	doi:10.1140/epjds/s13688-015-0055-z.
	
	\bibitem{monechi2015congestion}
	Monechi B, Servedio VD, Loreto V.
	\newblock {Congestion transition in air traffic networks}.
	\newblock PloS one. 2015;10(5):e0125546.
	
	\bibitem{sen2003small}
	Sen P, Dasgupta S, Chatterjee A, Sreeram P, Mukherjee G, Manna S.
	\newblock {Small-world properties of the Indian railway network}.
	\newblock Physical Review E. 2003;67(3):036106.
	
	\bibitem{guimera2005worldwide}
	Guimera R, Mossa S, Turtschi A, Amaral LN.
	\newblock {The worldwide air transportation network: Anomalous centrality,
		community structure, and cities' global roles}.
	\newblock Proceedings of the National Academy of Sciences.
	2005;102(22):7794--7799.
	
	\bibitem{WorldUrbanization_2014}
	{United Nations Secretariat}.
	\newblock {World Urbanization Prospects: The 2014 Revision}.
	\newblock Department of Economic and Social Affairs; 2014.
	
	\bibitem{galton1881construction}
	Galton F.
	\newblock {On the construction of isochronic passage-charts}.
	\newblock In: Proceedings of the Royal Geographical Society and Monthly Record
	of Geography. vol.~3. JSTOR; 1881. p. 657--658.
	
	\bibitem{zastrow2015data}
	Zastrow M.
	\newblock Data visualization: Science on the map.
	\newblock Nature News. 2015;519(7541):119.
	
	\bibitem{Banavar1999}
	Banavar JR, Maritan A, Rinaldo A.
	\newblock {Size and form in efficient transportation networks}.
	\newblock Nature. 1999 may;399(6732):130--132.
	\newblock Available from: \url{http://www.nature.com/articles/20144}.
	doi:10.1038/20144.
	
	\bibitem{Louf2014}
	Louf R, Roth C, Barthelemy M.
	\newblock {Scaling in Transportation Networks}.
	\newblock PLoS ONE. 2014 jul;9(7):e102007.
	\newblock Available from:
	\url{http://dx.plos.org/10.1371/journal.pone.0102007}.
	doi:10.1371/journal.pone.0102007.
	
	\bibitem{Masucci2015}
	Masucci AP, Arcaute E, Hatna E, Stanilov K, Batty M.
	\newblock {On the problem of boundaries and scaling for urban street networks.}
	\newblock Journal of the Royal Society, Interface. 2015 oct;12(111):20150763.
	\newblock Available from: \url{http://www.ncbi.nlm.nih.gov/pubmed/26468071
		http://www.pubmedcentral.nih.gov/articlerender.fcgi?artid=PMC4614511}.
	doi:10.1098/rsif.2015.0763.
	
	\bibitem{Arcaute2016}
	Arcaute E, Molinero C, Hatna E, Murcio R, Vargas-Ruiz C, Masucci AP, et~al.
	\newblock {Cities and regions in Britain through hierarchical percolation}.
	\newblock Royal Society Open Science. 2016 apr;3(4):150691.
	\newblock Available from:
	\url{http://rsos.royalsocietypublishing.org/lookup/doi/10.1098/rsos.150691}.
	doi:10.1098/rsos.150691.
	
	\bibitem{Li2017}
	Li R, Dong L, Zhang J, Wang X, Wang WX, Di Z, et~al.
	\newblock {Simple spatial scaling rules behind complex cities}.
	\newblock Nature Communications. 2017 dec;8(1):1841.
	\newblock Available from:
	\url{http://www.nature.com/articles/s41467-017-01882-w}.
	doi:10.1038/s41467-017-01882-w.
	
	\bibitem{boussauw2018planning}
	Boussauw K, Van~Meeteren M, Sansen J, Meijers E, Storme T, Louw E, et~al.
	\newblock Planning for agglomeration economies in a polycentric region:
	Envisioning an efficient metropolitan core area in Flanders. 2018;.
	
	\bibitem{hansen1959accessibility}
	Hansen WG.
	\newblock {How accessibility shapes land use}.
	\newblock Journal of the American Institute of planners. 1959;25(2):73--76.
	
	\bibitem{black1977accessibility}
	Black J, Conroy M.
	\newblock {Accessibility measures and the social evaluation of urban
		structure}.
	\newblock Environment and Planning A. 1977;9(9):1013--1031.
	
	\bibitem{paez2012measuring}
	P{\'a}ez A, Scott DM, Morency C.
	\newblock {Measuring accessibility: positive and normative implementations of
		various accessibility indicators}.
	\newblock Journal of Transport Geography. 2012;25:141--153.
	
	\bibitem{bok2016comparable}
	Bok J, Kwon Y.
	\newblock {Comparable Measures of Accessibility to Public Transport Using the
		General Transit Feed Specification}.
	\newblock Sustainability. 2016;8(3):224.
	
	\bibitem{schmocker2008mode}
	Schm{\"o}cker JD, Quddus MA, Noland RB, Bell MG.
	\newblock {Mode choice of older and disabled people: a case study of shopping
		trips in London}.
	\newblock Journal of Transport Geography. 2008;16(4):257--267.
	
	\bibitem{miller1999measuring}
	Miller HJ.
	\newblock {Measuring space-time accessibility benefits within transportation
		networks: basic theory and computational procedures}.
	\newblock Geographical analysis. 1999;31(1):1--26.
	
	\bibitem{geurs2004accessibility}
	Geurs KT, Van~Wee B.
	\newblock {Accessibility evaluation of land-use and transport strategies:
		review and research directions}.
	\newblock Journal of Transport geography. 2004;12(2):127--140.
	
	\bibitem{weiss2018global}
	Weiss D, Nelson A, Gibson H, Temperley W, Peedell S, Lieber A, et~al.
	\newblock A global map of travel time to cities to assess inequalities in
	accessibility in 2015.
	\newblock Nature. 2018;553(7688):333.
	
	\bibitem{Forbes_1964}
	Forbes J.
	\newblock {Mapping accessibility}.
	\newblock Scottish Geographical Magazine. 1964 apr;80(1):12--21.
	\newblock doi:10.1080/00369226408735915.
	
	\bibitem{luxen-vetter-2011}
	Luxen D, Vetter C.
	\newblock {Real-time routing with OpenStreetMap data}.
	\newblock In: Proceedings of the 19th ACM SIGSPATIAL International Conference
	on Advances in Geographic Information Systems. GIS '11. New York, NY, USA:
	ACM; 2011. p. 513--516.
	\newblock Available from: \url{http://doi.acm.org/10.1145/2093973.2094062}.
	doi:10.1145/2093973.2094062.
	
	\bibitem{OpenStreetMap}
	OpenStreetMap;.
	\newblock \textit{Available at} \url{https://www.openstreetmap.org/}.
	
	\bibitem{gtfs_provider}
	{TransitFeeds - Public transit feeds from around the world};.
	\newblock Accessed on November 2017.
	\newblock https://transitfeeds.com/.
	\newblock Available from: \url{https://transitfeeds.com/}.
	
	\bibitem{OECDCities}
	OECD.
	\newblock Redefining Urban.
	\newblock OECD Publishing; 2012.
	\newblock Available from:
	\url{https://www.oecd-ilibrary.org/content/publication/9789264174108-en}.
	doi:https://doi.org/https://doi.org/10.1787/9789264174108-en.
	
	\bibitem{popGridEurostat}
	Eurostat population grid;.
	\newblock \textit{Available at}
	\url{http://ec.europa.eu/eurostat/web/gisco/geodata/reference-data/population-distribution-demography/geostat}.
	
	\bibitem{Eurostat_2017}
	{Statistics of European Cities};.
	\newblock \textit{Available at}
	\url{http://ec.europa.eu/eurostat/statistics-explained/index.php/Statistics_on_European_cities}.
	
	\bibitem{sedacV4}
	for International Earth Science Information Network CIESIN Columbia~University
	C. {Gridded Population of the World, Version 4 (GPWv4): Population Count}.
	\newblock Palisades, NY: NASA Socioeconomic Data and Applications Center
	(SEDAC); 2016.
	\newblock Available from: \url{http://dx.doi.org/10.7927/H4X63JVC}.
	
	\bibitem{delling2009engineering}
	Delling D, Sanders P, Schultes D, Wagner D.
	\newblock {Engineering route planning algorithms}.
	\newblock In: Algorithmics of large and complex networks. Springer; 2009. p.
	117--139.
	
	\bibitem{disser2008multi}
	Disser Y, M{\"u}ller-Hannemann M, Schnee M.
	\newblock {Multi-criteria shortest paths in time-dependent train networks}.
	\newblock In: International Workshop on Experimental and Efficient Algorithms.
	Springer; 2008. p. 347--361.
	
	\bibitem{Accessibility_Houston}
	Akhavan A, Phillips NE, Du J, Chen J, Sadeghinasr B, Wang Q. 
	\newblock {Accessibility Inequality in Houston}.
	\newblock In: IEEE Sensors Letters. 2019;3(1):1–4.

	
	\bibitem{dibbelt2013intriguingly}
	Dibbelt J, Pajor T, Strasser B, Wagner D.
	\newblock {Intriguingly simple and fast transit routing}.
	\newblock In: International Symposium on Experimental Algorithms. Springer;
	2013. p. 43--54.
	
	\bibitem{yan2014universal}
	Xiao-Yong Y., Chen Z., Ying F., Zengru D., Wen-Xu W.
	\newblock {Universal predictability of mobility patterns in cities}.
	\newblock In: Journal of The Royal Society Interface. 2014. p. 20140834
	
	\bibitem{kolbl2003energy}
	K{\"o}lbl R, Helbing D.
	\newblock {Energy laws in human travel behaviour}.
	\newblock New Journal of Physics. 2003;5(1):48.
	
	\bibitem{clark2000monocentric}
	Clark WA.
	\newblock Monocentric to polycentric: New urban forms and old paradigms.
	\newblock A Companion to the City. 2000;p. 141--154.
	
	\bibitem{clark_1951}
	Clark C.
	\newblock {Urban population densities}.
	\newblock Journal of the Royal Statistical Society. 1951;114:490--496.
	
	\bibitem{muth_1969}
	Muth R.
	\newblock {Cities and Housing}.
	\newblock University of Chicago Press, Chicago; 1969.
	
	\bibitem{mills_1970}
	Mills ES.
	\newblock {Urban density functions}.
	\newblock Urban Studies. 1970;7:5--20.
	
	\bibitem{db1}
	Biazzo I, Monechi B, Loreto V. public-transport-analysis; 2018.
	\newblock Open source code,
	\url{https://github.com/CityChrone/public-transport-analysis}.
	\newblock doi:10.5281/zenodo.1309835.
	
	\bibitem{db2}
	Biazzo I, Monechi B, Loreto V. openData; 2018.
	\newblock Open data available at \url{https://github.com/CityChrone/openData}.
	\newblock doi:10.5281/zenodo.1309927.
	
	\bibitem{Bast_2016}
    Bast H, Delling D, Goldberg A, Müller-Hannemann M, Pajor T, Sanders P, et~al.
    \newblock {Route Planning in Transportation Networks}.
    \newblock In: Algorithm Engineering. Springer International Publishing; 2016.
      p. 19--80.
    \newblock Available from: \url{https://doi.org/10.1007%2F978-3-319-49487-6_2}.
    
    \bibitem{delling2014round}
    Delling D, Pajor T, Werneck RF.
    \newblock {Round-based public transit routing}.
    \newblock Transportation Science. 2014;49(3):591--604.

\end{thebibliography}

\appendix

\section*{Appendix}

\section*{Robustness of the accessibility metrics definitions respect travel time distributions}

In the definitions of the accessibility metrics, see for instance eq.(3) in the main text, we use a travel time distribution (TTD) $f(\tau)$ in order to average over time. The choice of the distribution used to average travel times is somehow arbitrary. 

\begin{figure}[!t]
	\includegraphics[angle=-90, width=0.95\textwidth]{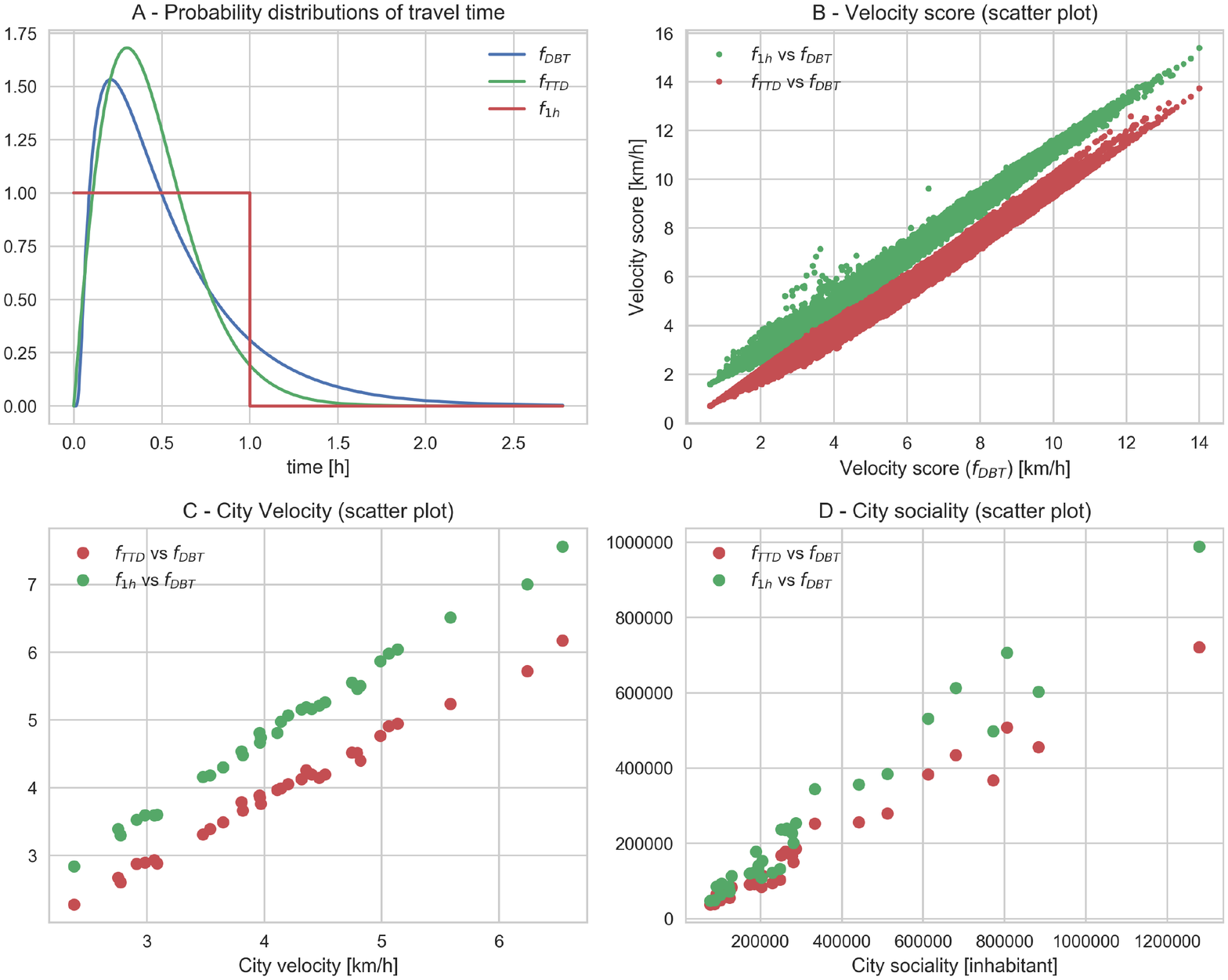}
	\caption{{\bf Robustness of the accessibility measure definition.} {\bf Panel A.} Plots of the three different travel time probability distributions. The curve $f_{DBT}(t)$ is taken from a survey on the daily use of public transports\cite{kolbl2003energy}, the $f_{TTD}$ is extracted from Oyster card journeys of London \cite{Gallotti2015,Gallotti2016}, and the $f_{1h}$ is a flat distribution between $0$ and $1$ hour. {\bf Panel B.} Scatter plot of the velocity score computed in all cities with three different travel time distributions. On x-axis is the value computed with the $f_{DBT}(t)$ distribution. {\bf Panel C(D).} Scatter plot of the city velocity(city sociality) computed with three different travel time distributions. On x-axis there is the values computed with the $f_{DBT}(t)$ distribution.}
	\label{fig:plots_robustness}
\end{figure}

In this section, we show that the accessibility metrics computed with differerent TTD  are highly correlated with one another. Hence, different choices of TTD lead to qualitatively similar results. The first TTD considered is derived from fits of an empirical daily budget time distribution based on surveys about times spent on a bus by UK citizen \cite{kolbl2003energy}. From the daily budget distribution, we obtain the travel time distribution considering that on average the people perform (with a very rough approximation) two trip per day. The empirical law has been derived in \cite{kolbl2003energy}, and it is the following:
\begin{equation}
f_{DBT}(t) = N * \exp( -\alpha T_{bus}/ t - \beta t / T_{Bus}) 
\label{eq:helb}
\end{equation}
\noindent { where $N$ is a normalization constant that ensure that $\int_{0}^{\infty}f_{DBT}(t)dt = 1$.} Then $\alpha = 0.2$, $\beta = 0.7$ and $T_{bus} = 67\;$min are obtained by the best fit on surveys data about public transport habits. 
Then from the daily budget distribution eq.\ref{eq:helb} we obtain the travel time distribution computing it a $2t$, $f_{DBT}(\tau) = f_{DBT}(2t)$, where we are considering that the daily budget time is spent in two trips. This is the distribution used to compute all the quantities shown in the main text. Then we consider the travel time distribution ($f_{TTD}$) extracted from Oyster card journeys on bus, Tube, Docklands Light Railway and London Overground \cite{Gallotti2015,Gallotti2016} and a normalized flat distribution $f_{1h}$, between $0$ and $1$ hour. The distributions are shown in fig.\ref{fig:plots_robustness}(panel - A).
The fig.\ref{fig:plots_robustness}(panel B) shown the velocity score of all points of all cities in our data-set computed with the $f_{TTD}$ (red) and $f_{1h}$ (green) distributions as a function of the velocity score computed with the distribution $f_{DBT}$ used in the main text.
The plot shows the high correlation between the velocity scores computed with these three different TTD.
Then we check how the global quantity as the city velocity and the city sociality used to rank the city are only slightly affected by the choice of the travel time distribution. In  Fig.\ref{fig:plots_robustness}, panel C (panel D) it is shown the scatter plot of the city velocity (city sociality) computed with the $f_{TTD}$ (red) and $f_{1h}$ (green) distributions as a function of the city velocity computed with the distribution $f_{DBT}$.
Also, in this case, the correlations between values are very high keeping in most cases the same rank order and relative distances between cities.

\section*{Fast and efficient routing algorithm for public transport networks}

The accessibility measures computed in the present work are based on the computations of the minimum travel time between each point in the cities using public transport. 
Due to the fact that the number of minimum travel times to be computed is of the order $10^9$ for each city, using an efficient routing algorithm is indeed mandatory. 
Driving times over the road network can be computed efficiently in a few milliseconds or less at the continental scale, but this is not the case of travel times with public transport \cite{Bast_2016}. This is because the approaches and speedup techniques used for road network routing algorithms fail \cite{Bast_2016} or are not so effective on public transport networks. In the last years, different approaches have emerged in literature where the most promising ones are the RAPTOR algorithm \cite{delling2014round} and the CSA Algorithm \cite{dibbelt2013intriguingly}. Both these algorithms take a different approach by not looking at the graph structure of the problem. However, these algorithms have a significant limitation when considering footpaths between public transport stops in order to change means of transportation, reducing the performances in urban systems.
The algorithm we used is based on the CSA algorithm, but with an essential modification that allows to use it in urban systems considering realistic footpaths to move between stops on foot and change mean of transport. We first describe the CSA algorithms and then the modified one, the Intransitive Connection Scan Algorithm (ICSA).
\subsection*{Connections Scan Algorithm (CSA)}
A public transit network is defined by its timetable. A timetable consists of a set of stops, a set of connections and a set of footpaths. The stops are the points on the map where a traveler can enter or exit from a vehicle.  A \emph{connection} $c$ represents a vehicle departing from the stop $s_{dep}(c)$ at time $t_{dep}(c)$ and arriving at the stop $s_{arr}(c)$ at time $t_{arr}(c)$ without intermediate halt. 
Movements between two stops performed on foot are called \emph{footpaths} and are treated separately with respect to the connections.
Given the timetable it is possible to compute the time needed to reach all the other stops given a starting time $t_0$ at the stop $s_{start}$. We label each stop $s_i$ with its arrival time $\tau[s_i]$ and we set them all at starting to infinity $\tau[s_i] = \infty$ except for the starting stop $\tau[s_{start}] = t_0$ that we set to its starting time. Then we build an array containing the connections, ordered by their $t_{dep}(c)$.
A connection is defined as \emph{reachable} if the time $t_{dep}(c)$ of starting stop $s_{dep}(c)$ of the connection $c$ is equal or larger than the time $\tau[s_{dep}(c)]$ of the stop $s_{dep}(c)$.
The Connections Scan Algorithm (CSA) scans the ordered array of connections $\{c\}$, testing if each $c$ is \emph{reachable}. If $c$ can be reached and if the arrival time $\tau[s_{arr}(c)]$ is larger of the arrival time $t_{arr}(c)$ of the connection, the connection is \emph{relaxed}, meaning that the $\tau[s_{arr}(c)]$ is update to the earlier arrival time $t_{arr}(c)$. After the entire array of connections is scanned the labels $tau$ contain the earliest arrival time for each stop starting from $s_{start}$.
In the above description we do not handle footpaths. Hence, in order to consider them each time the algorithm relaxes a connection, it checks all the outgoing footpath $f[s_{arr}]$ of $s_{arr}(c)$ and updates the time of the neighbors accordingly. The algorithm requires footpaths to be \emph{closed under transitivity} to ensure correctness. This means that if there is a footpath from stop $s_{a}$ to stop $s_{b}$ and a footpath from the stop $s_{b}$ to the stop $s_{c}$ there must be a footpath between $s_{a}$ and $s_{c}$. So for every connected component of stops connected by footpaths we need all the footpaths connecting them. 
Since the number of footpaths grows quadratically with the number of stops, it is computationally infeasible to consider them all. In order to reduce computational time and yet considering realistic footpaths, we allowed up to $15$ minutes of walk between stops.
Despite with this choice all the stops of a considered city belong to the same connected component, the fact that some footpaths are not considered might lead to an overestimation of the minimum travel time between two locations due to the lack of closeness.
%
In the next section, we describe a new version of the CSA algorithms that solves this problem. A pseudo-code of the CSA algorithm is shown in Fig.~\ref{alg:CSA algorithm}

\begin{figure}
	\begin{algorithm}[H] 
		\lFor{all stops $s$}{$\tau[s]\gets \infty$}
		$\tau[s_{start}] \gets t_0$\;
		\BlankLine
		\For{all connections $c$ increasing by $t_{dep}(c)$}{
			\If{$\tau[s_{dep}(c)] \leq t_{dep}(c)$}{
				\If{$\tau[s_{arr}(c)] > t_{arr}(c)$}{
					$\tau[s_{arr}(c)] \gets t_{arr}(c)$\;
					\For{all footpaths $f$ from $s_{arr}(c)$}{
						$\tau[f_{arr}] \gets \min \{ \tau[f_{arr}], \tau[s_{arr}(c)] + f_{dur} \}$\; 
					}
				}
			}
		}
		
	\end{algorithm}
	
	\caption{Connection Scan algorithm.}
	\label{alg:CSA algorithm}
\end{figure}
\subsection*{Intransitive Connections Scan Algorithm (ICSA)}
\begin{figure}[htbp]
	\centerline{\includegraphics[angle=-90, width=3.5in]{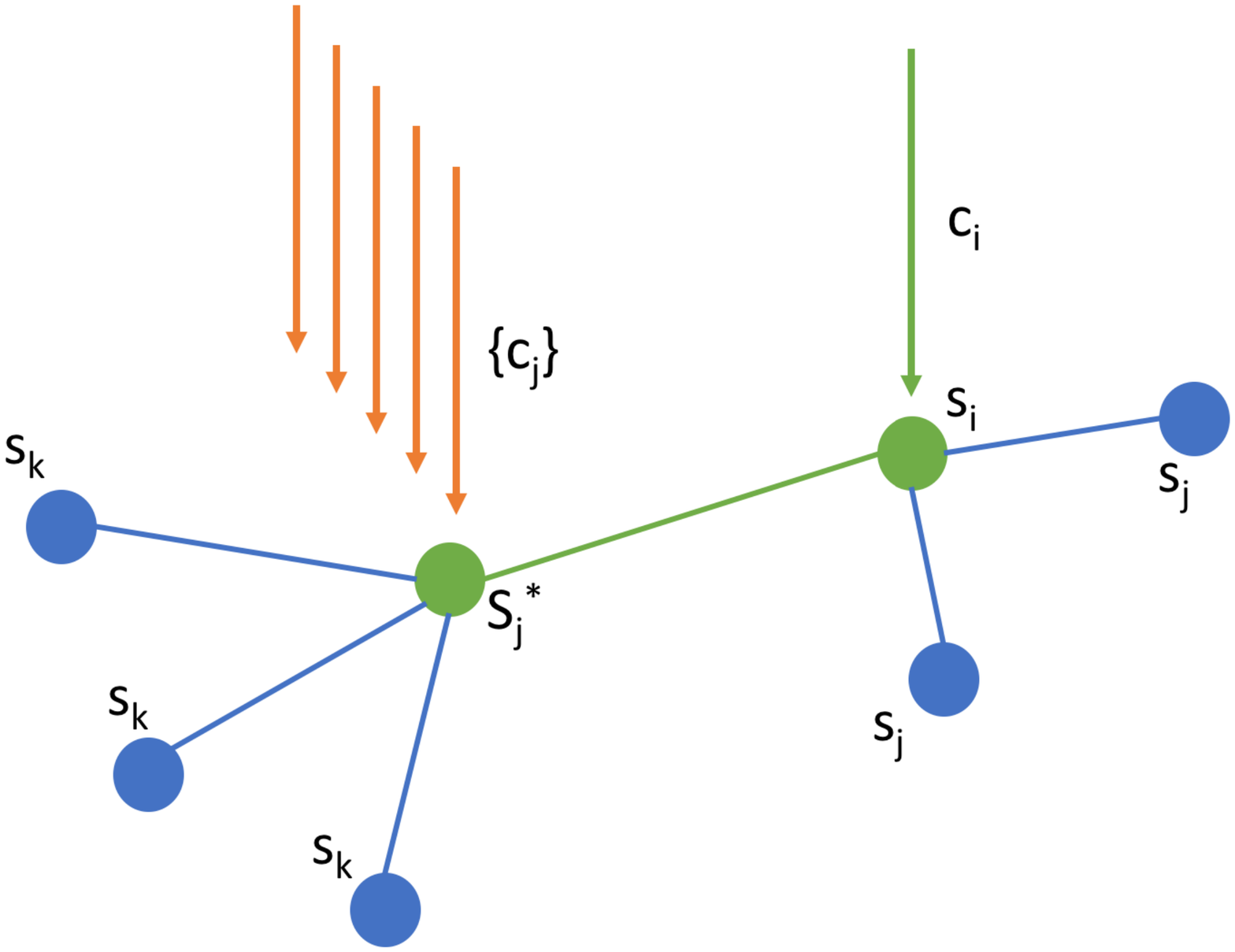}}
	\caption{In this cartoon shows the updating process of the CSA algorithm and the error when the closeness by transitivity of the footpaths between stops is not considered. First a connection $c$ arriving at the stop $s_i$ relaxes, i.e. update the arrival time $\tau[s_i]$ of $s_i$. Then the arrival time $\tau[s^*_{j}]$ of the neighbour stop $s^*_{j}$ is update through the footpath. Then other connections $\{c_j\}$ arriving to $s^*_{j}$ do not relax and the neighbour $\{s_{k}\}$ are never update by footpaths from  $s^*_{j}$ causing possible errors.}
	\label{fig:error_csa}
\end{figure}

The variant of the CSA we propose here can correctly solve the earliest arrival time problem on public transport network considering footpaths between stops without imposing the closeness under transitivity. 
Let us consider the case where closeness is not enforced and all the footpaths lasting more than $15\;$ minutes are removed. For each stop $s_i$ consider a subsets of stops $\{s_{j}\}$ reachable with these footpaths. 
Thus, the journeys computed by the CSA algorithm could be incorrect due to missing travel time updates that should have been performed through the removed footpaths.
%
Consider the case, we have just relaxed a connection $c$, i.e. the arrival time $\tau[s_{i}(c)]$ of the arrival stop $s_{i}(c)$ is updated, see Fig.\ref{fig:error_csa}. Then the stops $\{s_{j}\}$, reachable by footpaths from $s_{i}(c)$, are checked and the arrival time $\tau[s^*_{j}]$ of a neighbour stop $s^*_{j}$ is updated. $s^*_{j}$ is also connected by footpaths to other stops $\{s_{k}\}$ (see Fig.\ref{fig:error_csa}), which under closeness should have been updated through footpaths connecting them directly with $s_i$. However, despite they could be still updated through the footpaths starting from $s_j$, this does not happen because the updating of arrival time ends to the first set of neighbor stops.
In worst cases, it could also happen that the remaining connections that arrive on $s_{j}$ never relax, because all of them arrive after the time $\tau[s_{j}]$. Hence, no one of its neighbors will be updated through footpaths connecting them to $s_{j}$. However, there could be journeys passing through $s_{j}$ that do not update $\tau[s_{j}]$ directly, but they might update some of its neighbors $\{s_{k}\}$ through footpaths. The CSA algorithm is not able to consider this.
The Intransitive Connections Scan algorithms (ICSA) overcomes this problem with a small modification of the original CSA algorithm, without increasing the running time and preserving its simplicity. The key is to consider two labels $\tau$ and $\tau^{f}$ for the arrival time for each stops. One represents the arrival time to the stop after the relaxation of one connection and the other due to the updating of the arrival time by footpaths. 
The ICSA algorithm is the same as CSA except for three modifications:
\begin{enumerate}
	\item a connection is considered reachable if its starting time $t_{dep}(c)$ is larger or equal either to arrival time of the stops $\tau[s_{dep}(c)]$ due to connection updating or to the arrival time $\tau^f[s_{dep}(c)]$ due to the footpaths update.
	\item Both the CSA and ICSA update the arrival time of the stops in two ways: by direct relaxation of the connections or by footpaths. The ICSA algorithm updates the arrival time $\tau[s]$ of the stop $s$ when it is updated by connections, and the arrival time $\tau^f[s]$ when it is updated by footpaths. 
	\item After the complete scanning of the connections array, the arrival time taken for each stops $s$ is the best arrival time between the two labels $tau[s]$, $tau^{f}[s]$.
\end{enumerate}
A pseudo-code implementation scheme is shown in fig.\ref{alg:ICSA algorithm}.

\begin{figure}
	\begin{algorithm}[H]
		\lFor{all stops $s$}{$\tau[s]\gets \infty$}
		\lFor{all stops $s$}{$\tau^f[s]\gets \infty$}
		$\tau[s_{start}] \gets t_0$\;
		\BlankLine
		\For{all connections $c$ increasing by $t_{dep}(c)$}{
			\If{$\tau[s_{dep}(c)] \leq t_{dep}(c)$ or $\tau^f[s_{dep}(c)] \leq t_{dep}(c)$}{
				\If{$\tau[s_{arr}(c)] > t_{arr}(c)$}{
					$\tau[s_{arr}(c)] \gets t_{arr}(c)$\;
					\For{all footpaths $f$ from $s_{arr}(c)$}{
						$\tau^f[f_{arr}] \gets \min \{ \tau^f[f_{arr}], \tau[s_{arr}(c)] + f_{dur} \}$\; 
					}
				}
			}
		}
		\lFor{all stops $s$}{$\tau[s]\gets min(\tau[s], \tau^f[s])$}
	\end{algorithm}
	
	\caption{Intransitive Connection Scan Algorithm}
	\label{alg:ICSA algorithm}
\end{figure}
These modifications allow to correctly solve the problem of finding the earliest arrival time to stops, given a set of connections and footpaths connecting them, without the constraint of the closeness under transitivity of footpaths.

\begin{figure}[htbp!]
	\centerline{\includegraphics[width=0.9\textwidth]{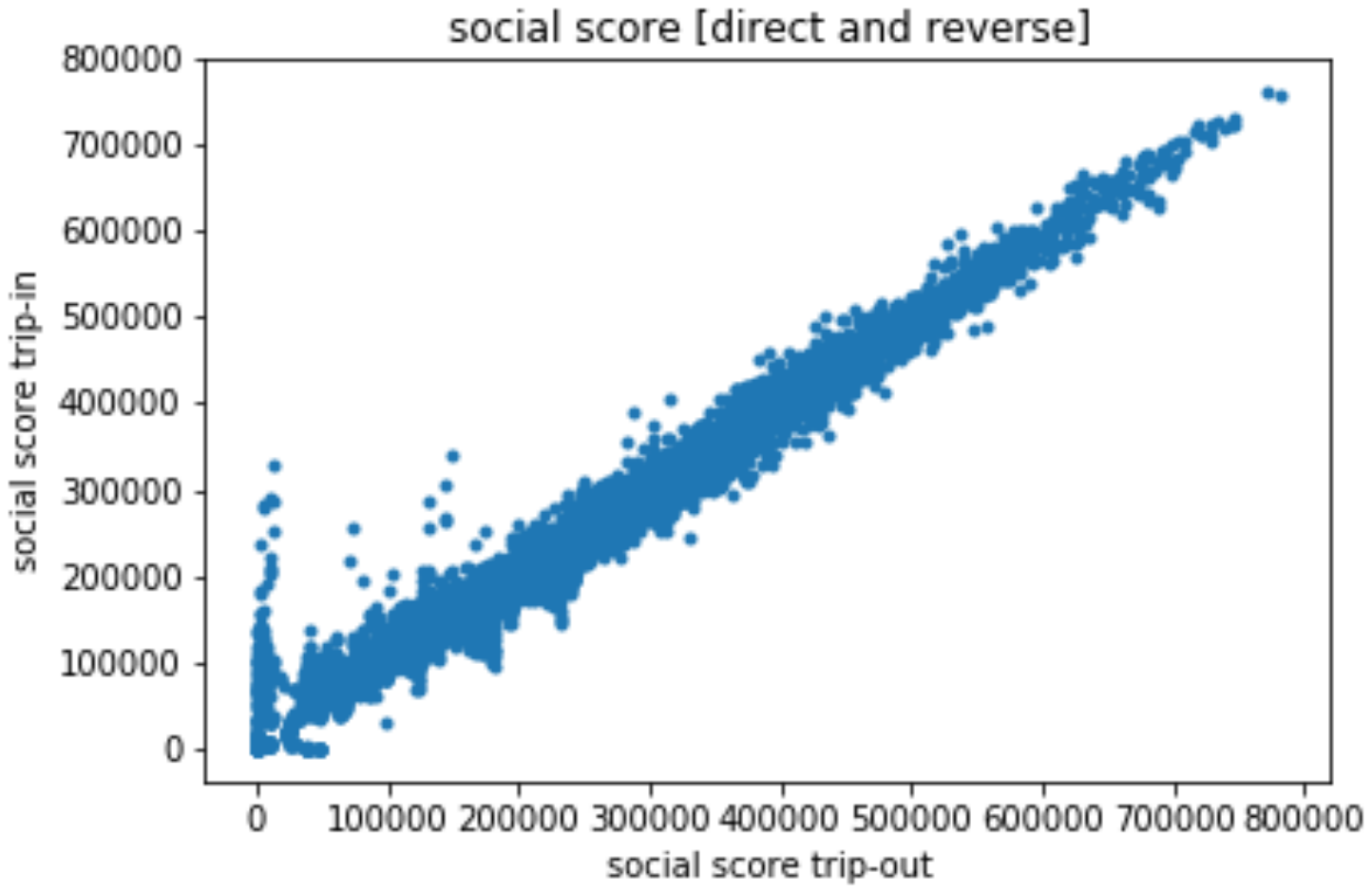}}
	\caption{{\bf Scatter plot of sociality score compute with travel time of incoming and out-coming trip.} The sociality score is based on the computation of the travel time of trips connecting the point considered to all the others. When considering the travel time of the out coming trips the sociality score represents the amount of people is possible to reach at home from the considered point. Instead when the travel time of incoming trips are taken into account the sociality score represent the amount of people can easily reach the considered points. The two quantity are very similar to each others, because, on average, the travel time trip of the out-coming and incoming trips are very similar. In the figure there is the scatter plot between the sociality score computed with incoming and outcoming trips for Rome.}
\end{figure}

\begin{figure}[htbp]
    \centerline{\includegraphics[angle=-90, width=1\textwidth]{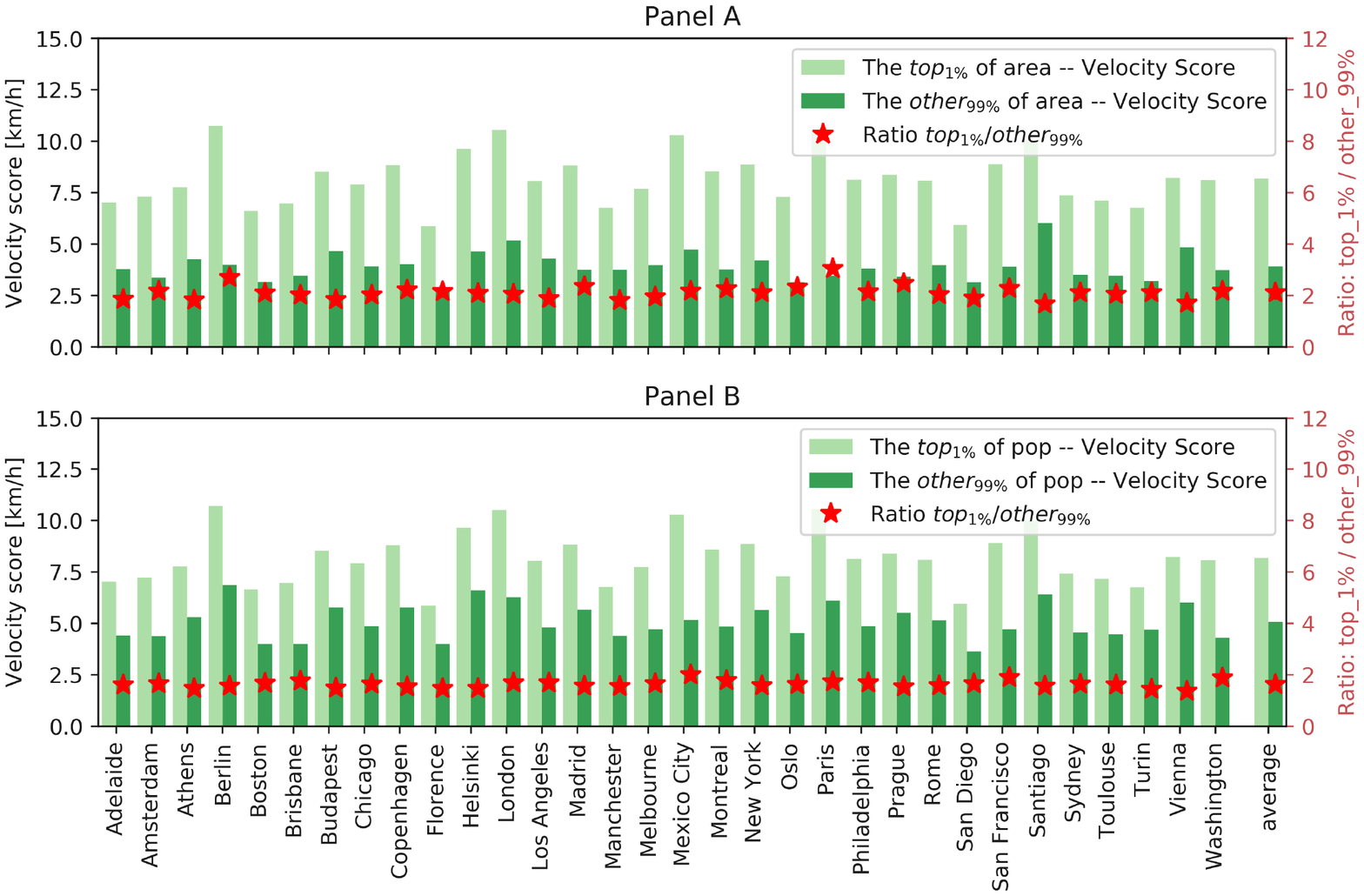}}
    \caption{{\bf Panel A}: Average values of the velocity scores among the hexagons featuring the top $1\%$ (light green) of the values of the velocity score as compared to the remaining $99\%$ (dark green) for all cities analyzed. The last columns report the same values averaged over all cities. Red stars mark, on the right $y$-axis,  are the ratios between the average values of top $1\%$ and the other $99\%$. {\bf Panel B}: Average values of the velocity scores among the population with the highest top $1\%$ (light green) of the values of the velocity score as compared to the remaining $99\%$ (dark green) for the six selected cities. The last columns report the same values averaged over all the six cities. Red stars mark, on the right $y$-axis,  are the ratios between the average values of top $1\%$ and the other $99\%$. }
    \label{fig:one_ninenine_vel}
\end{figure}

\begin{figure}[htbp]
	\centerline{\includegraphics[width=1\textwidth]{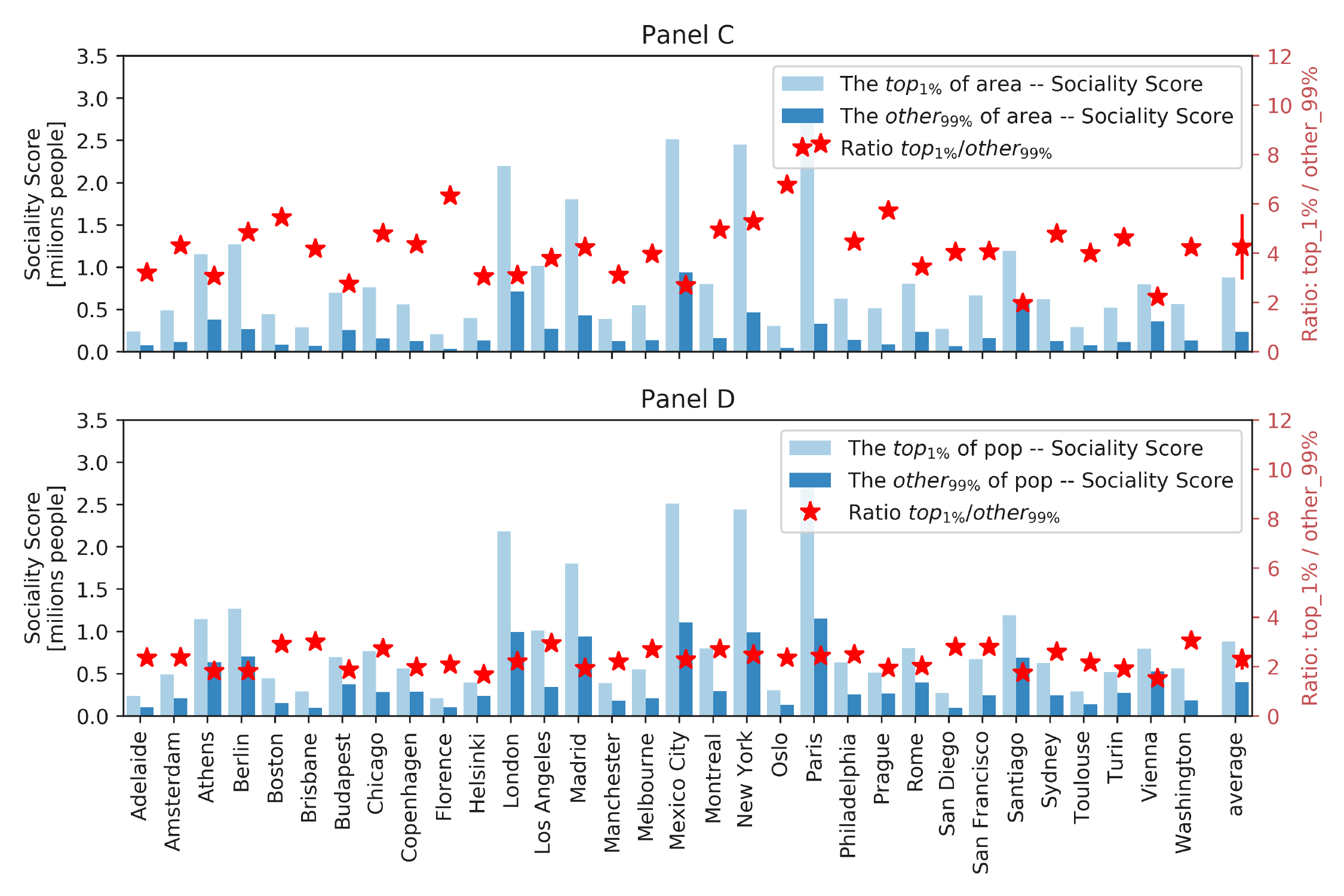}}
	\caption{{\bf Panel C}: Same as Panel A of Fig.\ref{fig:one_ninenine_vel} for the Sociality score. {\bf Panel D}: Same as Panel B of Fig.\ref{fig:one_ninenine_vel} for the Sociality score.}
\end{figure}

\begin{figure}[htbp]
	\centerline{\includegraphics[width=1\textwidth]{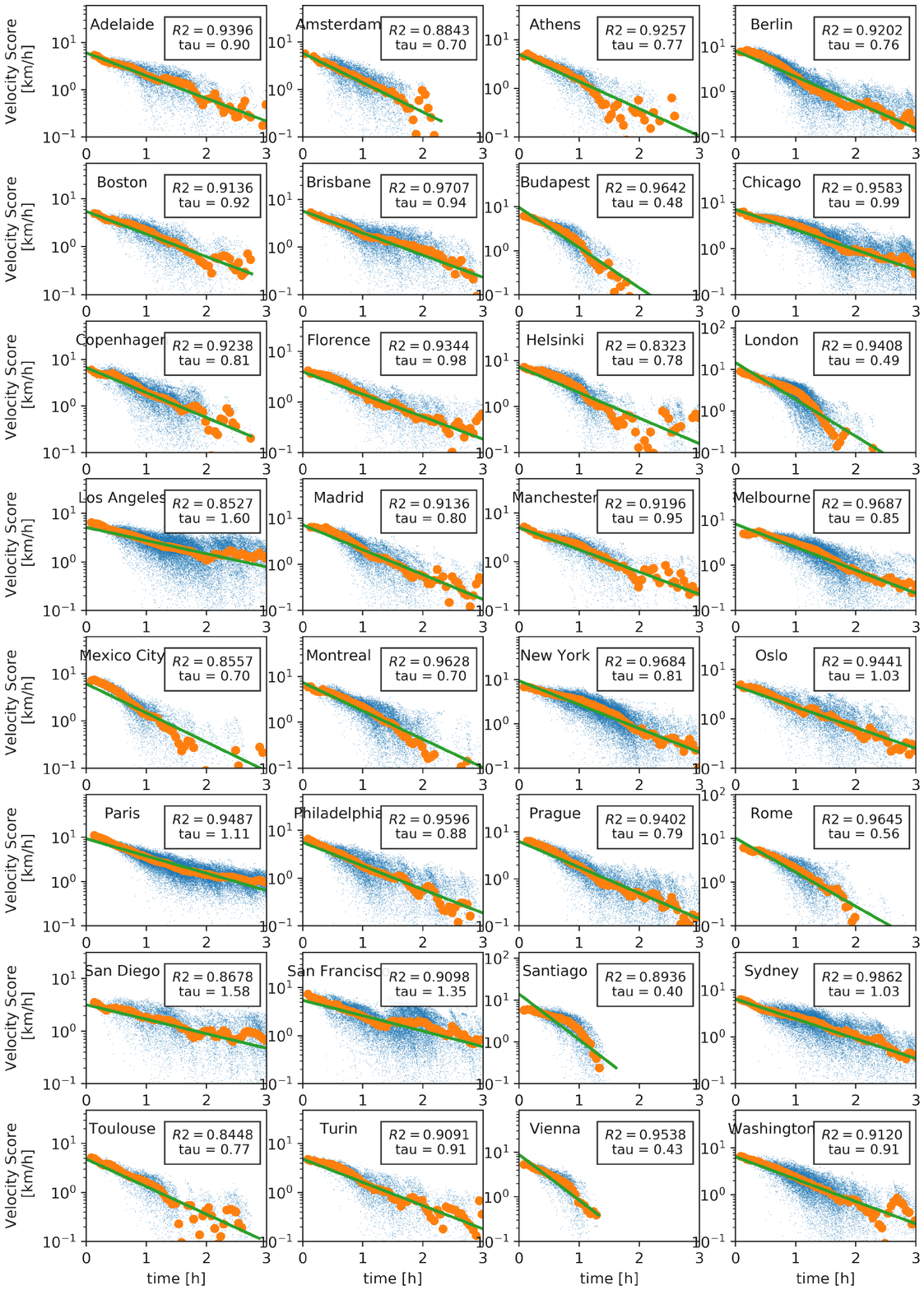}}
	\caption {{\bf Exponential decay of the velocity score with travel-times for the city center.} Velocity scores of hexagons at a given travel-time distance from the hexagon with the highest velocity score in each of the six selected city (blue dots). The orange points report a binning of the blue dots. The green line is the best fit of the data with the function 8 in the main text.}
\end{figure}
\begin{figure}[htbp]
	\centerline{\includegraphics[width=1\textwidth]{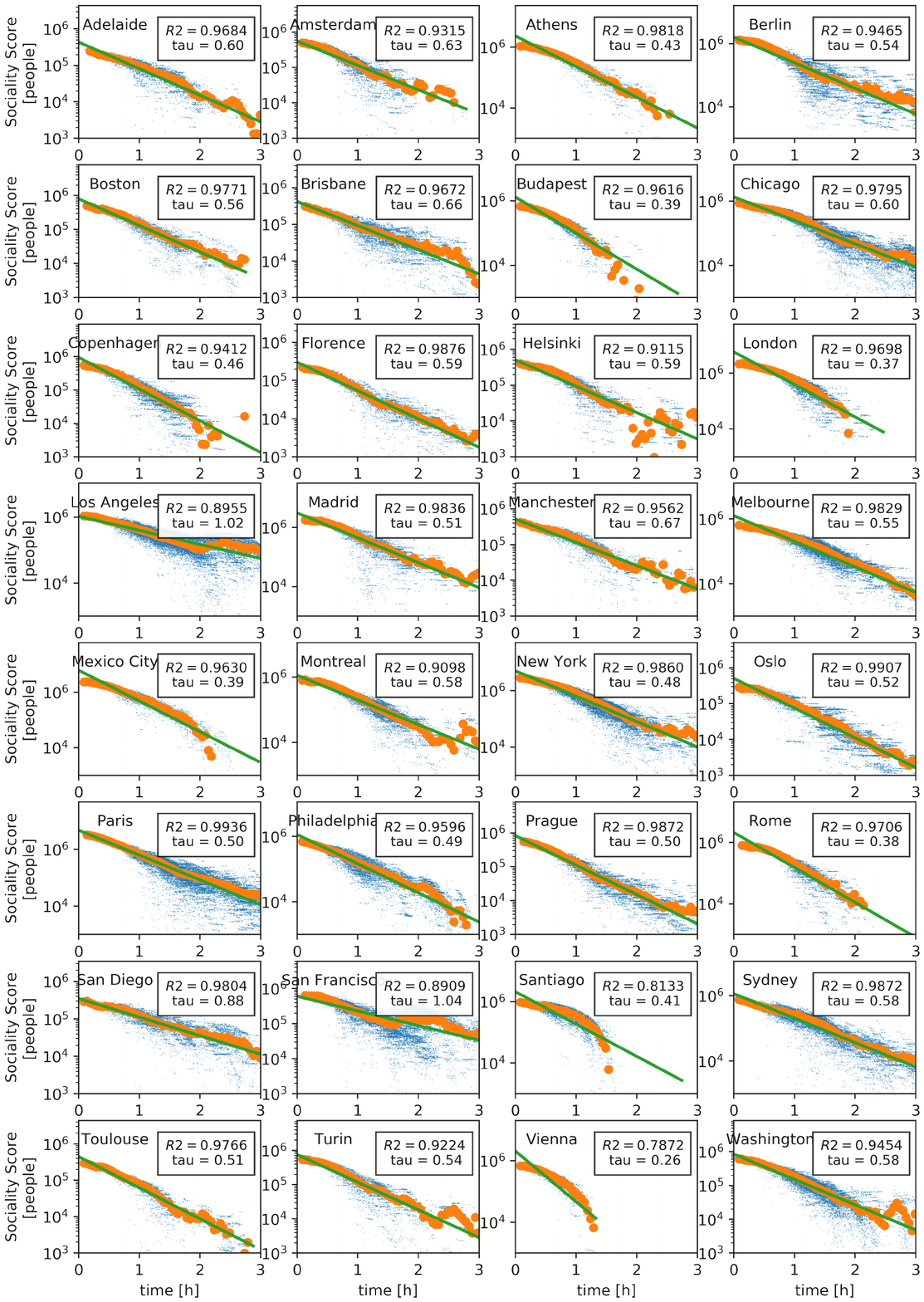}}
	\caption{{\bf Exponential decay of the sociality score with travel-times for the city center.} Sociality scores of hexagons at a given travel-time distance from the hexagon with the highest sociality score in each of the six selected city (blue dots). The orange points report a binning of the blue dots. The green line is the best fit of the data with the function 8 in the main text.}
\end{figure}

\end{document}